\documentclass{aa}  
\usepackage{graphicx}       
\usepackage{amsmath}        
\usepackage{xspace}         
\usepackage{algorithm}      
\usepackage{algpseudocode}  
\usepackage{txfonts}
\usepackage{xcolor}
\usepackage{comment}  
\usepackage[dvipsnames]{xcolor}
\usepackage[normalem]{ulem}

\newcommand{\Ms}{M$_{\odot}$\xspace}
\newcommand{\Zs}{Z$_{\odot}$\xspace}

\newcommand{\ex}{\textit{exchanged}\xspace}
\newcommand{\og}{\textit{original}\xspace} 
\newcommand{\be}{\textit{penetration factor}\xspace} 

\usepackage[normalem]{ulem}

\newcommand{\pt}{\texttt{PETAR}\xspace}
\newcommand{\fdps}{\texttt{FDPS}\xspace}
\newcommand{\sdar}{\texttt{SDAR}\xspace}
\newcommand{\bse}{\texttt{BSE}\xspace}

\usepackage[colorlinks=true, linkcolor=blue, citecolor=blue, urlcolor=blue]{hyperref}

\begin{document} 

\title{Micro-Tidal Disruption Events in Young Star Clusters}

\author{
Sara Rastello\inst{1,2}\thanks{e-mail: sara.rastello@icc.ub.edu}
\and
Giuliano Iorio\inst{2}\thanks{e-mail: giuliano.iorio@icc.ub.edu}
\and
Mark Gieles\inst{2,3,4}
\and
Long Wang\inst{5,6}
}

\institute{
Departament de Física Quàntica i Astrofísica (FQA), Universitat de Barcelona (UB), c. Martí i Franquès 1, 08028 Barcelona, Spain
\and
Institut de Ciències del Cosmos (ICCUB), Universitat de Barcelona (UB), c. Martí i Franquès 1, 08028 Barcelona, Spain
\and
ICREA, Pg. Lluís Companys 23, 08010 Barcelona, Spain
\and
Institut d’Estudis Espacials de Catalunya (IEEC), Edifici RDIT, Campus UPC, 08860 Castelldefels (Barcelona), Spain
\and
School of Physics and Astronomy, Sun Yat-sen University, Daxue Road, Zhuhai, 519082, China
\and
CSST Science Center for the Guangdong-Hong Kong-Macau Greater Bay Area, Zhuhai, 519082, China
}
   \date{Received August 1, 2025; accepted XX XX, XXXX}

 
  \abstract
   {Dense young star clusters (YSCs) are ideal environments for dynamical interactions between stars and stellar compact objects, such as black holes and neutron stars. In such dense environments, stars can undergo close encounters with black holes and fall within their tidal radius, resulting in tidal disruption. These events, known as micro-tidal disruption events (micro-TDEs), are transient phenomena with potential multi-messenger signatures.}
   {We aim to quantify the nature, occurrence, and observational relevance of micro-TDEs across a wide range of cluster masses, densities, and metallicities, through an extensive exploration of the parameter space.}
   {We performed a suite of direct $N$-body simulations using the \texttt{PETAR} code, to which we implemented new prescriptions for modeling micro-TDEs. We constructed a set of realistic YSC models including primordial binaries, based on the observed Milky Way population. Our simulations incorporate stellar and binary evolution, supernova kicks, and stellar winds using the \bse code, and account for the Galactic tidal field via the \texttt{GALPY} library.}
   {We identify three primary dynamical channels for micro-TDE production: single star-single black holes  encounters, binary-mediated interactions (including supernova-kick triggers), and interactions involving higher-order multiple systems such as hierarchical triples and quadruples, as well as chaotic few-body interactions with more than three objects. Multiple encounters are the most efficient production channel which dominates the total rate: $\sim 350$-$450$~Gpc$^{-3}$~yr$^{-1}$. Micro-TDEs from YSCs are expected to be detectable by upcoming surveys, particularly the Legacy Survey of Space and Time, with detection rates potentially up to hundreds per year. The gravitational wave signals expected from micro-TDE peaks in the deci-Hertz band, making them accessible to future instruments such as the Lunar Gravitational Wave Antenna and the Deci-Hertz Interferometer Gravitational wave Observatory.
}
   {Micro-TDEs emerge as promising multi–messenger sources, potentially offering unique insights into star cluster dynamics, stellar collisions, and the population of dormant stellar-mass black holes, through both electromagnetic and gravitational wave observations.}

   \maketitle
%

\section{Introduction}
Tidal Disruption Events (TDEs) are transient phenomena that occur when a star is disrupted during a close pericenter passage near a super massive black hole (SMBH, see e.g. \citealt{hills1975}). 
The characteristic distance from the SMBH where a TDE occurs is the \textit{tidal radius}: 
\begin{equation}
    \centering
    r_{\rm t} \simeq r_* \left( \frac{m_{\rm BH}}{m_*} \right)^{1/3},
    \label{eq:td}
\end{equation}
which depends on the mass and the radius of the star 
($m_{\rm{*}}$ and $r_{\rm{*}}$, respectively), and on the black hole mass ($m_{\rm{BH}}$). 
When the pericenter of the encounter $r_{\rm{p}}$ falls within the tidal radius, \begin{equation}
        \centering
        r_{\rm{p}}\, \lesssim \, r_{\rm{t}},
	\label{eq:cond}
\end{equation}
the SMBH's tidal forces exceed the star’s self-gravity, leading to its partial or total disruption. Following the disruption, approximately half of the stellar debris becomes unbound, while the remaining bound material returns to pericentre on highly eccentric orbits and eventually forms an accretion disk around the SMBH \citep{Rees1988, Guillochon2013, Komossa2015}. The accretion of the stars's debris can power a luminous flare detectable across multiple electromagnetic (EM) wavelengths (X-ray, optical, UV) thanks to which hundreds of TDEs have  been identified so far \citep{Gezari2021, Hammerstein2023}.
In addition to their EM signatures, TDEs are also expected to produce bursts of gravitational waves (GWs) when the star is torn apart (see e.g. \citealt{Kobayashi2004, Toscani2020,Pfister2022,wevers2023}).

TDEs are not exclusive to SMBHs; stars can also be tidally disrupted by stellar-mass compact objects (COs), such as stellar black holes (BHs) and neutron stars (NSs). These events, referred to as "\textit{micro-TDEs}"\footnote{Through this paper TDEs refer to events involving SMBHs; those involving stellar-COs are referred to as micro-TDEs.}, a term first introduced by \citet{perets2016}, represent a scaled-down analogue of classical TDEs involving SMBHs.
Micro-TDEs
are expected to occur primarily in dense stellar environments such as young star clusters (YSCs), globular clusters (GCs), and nuclear star clusters (NSCs), where dynamical close encounters between stars and COs, are common. In these dense and massive star clusters (SCs), COs can be efficiently retained, facilitating interactions with stars \citep{giesers2018, kremer2019, Gieles2021,Rastello20,rastello2021,torniamenti2022,torniamenti2023,mas24}.

Previous analytical estimates \citep{perets2016} predicted that micro-TDEs occur in GCs at rates of approximately $3-10$ Gpc$^{-3}$ yr$^{-1}$ in the local Universe. \citet{kremer2019, kremer2021} estimated even higher rates for GCs and YSCs ($10-100$ Gpc$^{-3}$ yr$^{-1}$). Additionally, micro-TDEs are also expected to occur in NSC at rates up to $\sim 10$ Gpc$^{-3}$ yr$^{-1}$ \citep{Fragione2021} and in the disc of active galactic nucleus (AGN) at even higher rate $\approx 170$ Gpc$^{-3}$ yr$^{-1}$ \citep{Fragione2021,Yang2022,Li2025}.

A key difference between micro-TDEs and classical TDEs is that the former may also involve few-body systems and/or multiple COs, such as stellar binaries or binary compact objects (BCOs), including binary black holes (BBHs). \citet{Ryu2022, Ryu2023a, Ryu2023b, Ryu2024} extensively studied close encounters between main sequence (MS) stars and BHs in multiple 3-body configurations, using both Smoothed Particle Hydrodynamics (SPH) and moving-mesh codes. \citet{Lopez2019} showed through hydrodynamic simulations that micro-TDEs on BBHs in a SCs can alter the intrinsic spins of the two BHs. \citet{Samsing2019} demonstrated that BH-star disruptions by BBHs can be used to constrain the orbital period of the binary, and to link such events to the corresponding merger rates. \citet{Fragione2019} showed that secular dynamics in hierarchical triple systems can also produce BH-star TDEs at non-negligible rates ($\sim 10^{-4}$ yr$^{-1}$). 
\citet{kremer2019,kremer2021} showed that in GCs and YSCs micro-TDEs occur several times more frequently in binary mediated encounters compared to single-single encounters. \citet{Rastello2018}, through direct $N$-body simulations, found that micro-TDE on BBHs occur in Open Cluster (OCs) at a rate of $\sim 3-30 $ Gpc$^{-3}$ yr$^{-1}$. 

Micro-TDEs are also relevant in the context of low-mass X-ray binaries (LMXBs) and the recently discovered population of dormant BHs \citep{giesers2018,shenar2022a,elbadry2023a,elbadry2023b,panuzzo}. 
Unlike LMXBs, where accretion from the companion star powers detectable X-ray emission \citep{Avakyan2023}, these dormant BHs are X-ray silent and are identified through precise astrometric and spectroscopic measurements of the motion of their luminous companions. Although most of these systems have been discovered in the field \citep{elbadry2023a,elbadry2023b,shenar2022b}, a dynamical origin in stellar clusters has not been excluded \citep{rastello2023,tanikawa2023bh,dicarlo2024}. Notably, the most massive BH currently known in the Milky Way (MW), Gaia BH3  \citep[$m_{\rm{BH}}\approx 33$ \Ms,][]{panuzzo} has recently been shown to be associated with the stellar stream of the disrupted GC ED2 \citep{balbinot}, suggesting a dynamical origin \citep{marinpina2024}. If such long-period BH-star binaries are dynamically perturbed, for example, through binary-single or binary-binary encounters in  SCs, their orbits may shrink or become highly eccentric, potentially driving the companion star close enough to the BH to trigger a micro-TDE. In this framework, dormant BHs could represent progenitors of future micro-TDEs in dense stellar environments. 

\citet{perets2016} proposed that micro-TDEs could explain highly energetic phenomena such as ultra long gamma-ray bursts. \citet{kremer2021} proposed stellar-mass BH TDEs as possible progenitor of fast blue optical luminous transients (FBOT), which are often observed in star-forming galaxies.
 \citet{kremer2023} performed SPH simulations of BH-MS stars showing that radiation reprocessed by disk winds can produce bright UV/optical transients (peak luminosities \( \sim 10^{41-44} \) erg s\(^{-1}\) that could be detected by ongoing and upcoming survey as  Zwicky Transient Facility (ZTF) at Palomar Observatory \citep{Bellm2019,graham2019},  Legacy Survey of Space and Time (LSST) at the Vera C. Rubin Observatory \citep{Ivezic2019}, and Ultraviolet Transient Astronomy Satellite (ULTRASAT) \citep{Sagiv2014,BenAmi2022,Shvartzvald2024}.\\
 Although micro-TDEs have yet to be directly observed, there are currently three potential candidates, though none have been definitively confirmed. Two of these candidates, ASASSN-15lh \citep{dong2016} and ZTF19aailpwl \citep{frederik2021} are associated with AGN-disk \citep{Yang2022}, while the third, AT 2022aedm \citep{nicholl2023}, with an elliptical galaxy. However, these candidates are too luminous compared to models of micro-TDEs light curves and further analysis is needed to confirm their nature.\\
In addition to EM surveys, a significant boost in GW detection is expected in the coming years thanks to instruments such as the Lunar Gravitational Wave Antenna (LGWA, \citealt{harms2021}), the Laser Interferometer Space Antenna (LISA, \citealt{pau2017}), and the Deci-Hertz Interferometer Gravitational wave Observatory (DECIGO, \citealt{sato2017}).\\
Micro-TDEs could serve as an alternative and complementary source for multi-messenger observations, offering valuable insights into the properties of BHs in SCs. These observations would complement data from GW interferometers and surveys, such as those identifying dormant BHs \citep{panuzzo}, which are expected to boost with the next Gaia data release \citep{breivik2017}. This is especially significant given the realization within the scientific community that the large degeneracy in stellar and binary evolutionary models, combined with uncertainties in star formation models, has hindered our ability to obtain stringent constraints from GW observations alone. Therefore, accurate modeling of micro-TDEs is both crucial and timely. \\
In this work, we present a comprehensive study of micro-TDEs involving stars and BHs and NSs in YSCs. We model these environments using state-of-the-art, high-precision direct $N$-body simulations that integrate updated stellar evolution and tidal disruption prescriptions. Our approach captures the full range of dynamical interactions capable of producing stellar disruptions, accounting for all relevant configurations in which stars are destroyed during close encounters with COs. By exploring the demographics and parameter space of micro-TDEs in realistic SC environments, our results offer valuable predictions for their occurrence rates and observable properties, filling a critical gap in our understanding of stellar dynamics and CO interactions in these dense environments.\\
In Sect. \ref{sec:ic} we present the numerical code used to perform the suite of simulations, while Sect. \ref{sec:nbody} describes the initial conditions. In Sect. \ref{sec:res} we show the results, including the formation channels of micro-TDEs, the production efficiency, and the estimated rates. Sect. \ref{sec:disc} is focused on the discussion, including a comparison with previous works, expected detectability, and implications for GW astronomy. Caveats are discussed in Sect. \ref{sec:cav}. In Sect. \ref{sec:conc} we summarize our main findings and present our conclusions. 
Appendix~\ref{app:petar} describes the modifications implemented in the code to include the TDE prescription. Appendix~\ref{app:eff} and ~\ref{app:rates} provides two additional tables reporting the production efficiency and detectability fractions of micro-TDEs.

\section{Numerical Methods}
\label{sec:meth}
\subsection{Star cluster models}
\label{sec:ic}
\begin{figure}
 	\includegraphics[width=0.45\textwidth]{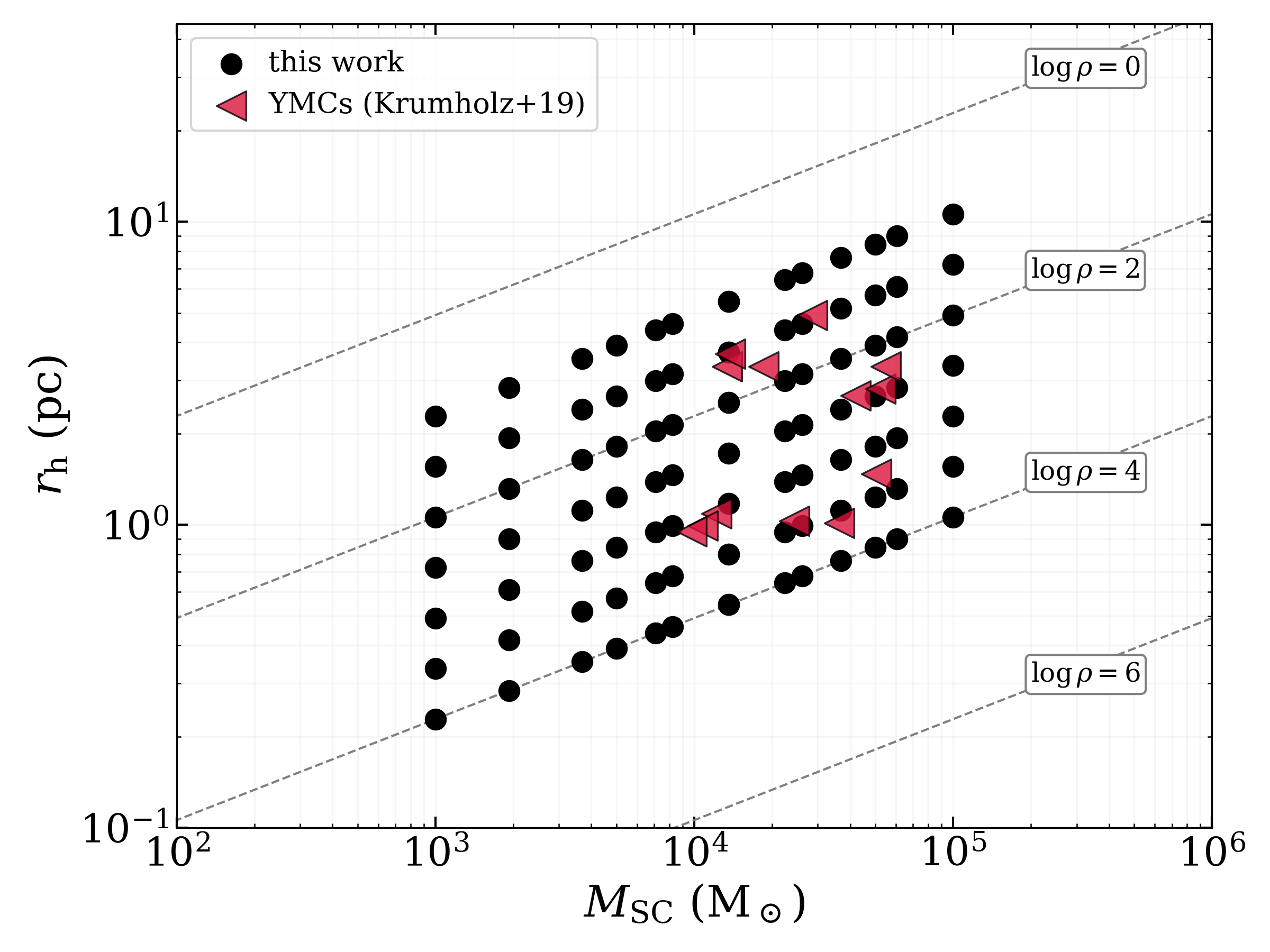}
   \caption{Grid of SCs models created in this work adapted from Fig. 9 in \cite{krumholz2019}. Black dots refers to combination of initial half-mass radius $r_{\rm h}$, cluster mass $M_{\rm SC}$ and initial density (at half mass radius) $\rho_{\rm h}$ used to construct the SC simulations. Red triangles refers to  observed YMCs \citep{krumholz2019}.}
    \label{fig:kru}
\end{figure}

We construct a grid of SC models to explore a wide range of initial conditions, varying the total initial cluster mass ($M_{\rm SC}$), the half-mass radius ($r_{\rm h}$) and corresponding half-mass density ($\rho_{\rm h}\equiv3M_{\rm SC}/(8\pi r_{\rm h}^3)$), based on the observed population of Galactic YSCs presented in \citet{krumholz2019}. A schematic representation of the parameter space explored is shown in Fig.~\ref{fig:kru}.

We sample the cluster mass in the range $10^3\,\mathrm{M}_\odot \leq M_{\mathrm{SC}} \leq 10^5\,\mathrm{M}_\odot$ using two overlapping logarithmically spaced grids. The first grid includes seven mass values in the range $10^3\,\mathrm{M}_\odot \leq M_{\mathrm{SC}} \leq 5 \times 10^4\,\mathrm{M}_\odot$,  while the second contains seven points in the range $5 \times 10^3\,\mathrm{M}_\odot \leq M_{\mathrm{SC}} \leq   10^5\,\mathrm{M}_\odot$.
For each cluster mass, we explore multiple values of $r_{\rm h}$, sampled on a logarithmic grid as well, thereby covering three orders of magnitude in $\rho_{\rm h}$.
To mitigate the impact of stochastic fluctuations in low-mass clusters, where the lower number of stars leads to greater variance, we fix the total simulated mass $M_\text{sim}$ per mass bin.
This strategy ensures uniform statistical sampling across the entire mass range and provides improved coverage of the region of parameter space occupied by observed Galactic YSCs in \citet{krumholz2019} (see Fig.~\ref{fig:kru}).

\begin{figure*}
 	  \centering
    \includegraphics[width=0.9\textwidth]{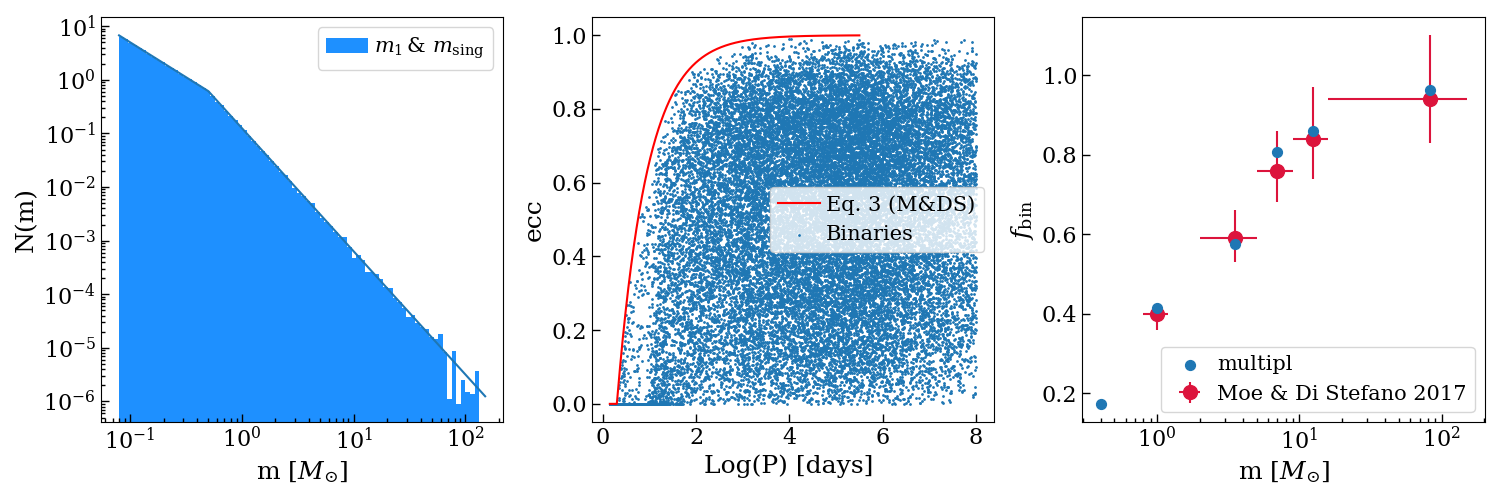}
   \caption {Properties of \og  binaries adopted in our simulations, shown for one representative cluster realization with initial mass $m_{\rm{SC}} \approx 9.8 \times 10^{4}$ \Ms and initial half mass radius $r_{\rm{h}} \approx 4$ pc.
\textit{Left panel}: mass distribution of primary stars ($m_{\rm{1}}$) in binaries and single stars ($m_{\rm{sing}}$), both sampled from a Kroupa IMF.
\textit{Central panel}: distribution of binary orbital eccentricities (ecc) as a function of period ($\log (\rm{P})$ in days). The red curve marks the exclusion boundary (Eq. 3 in \citealt{moe17}), to avoid Roche lobe filling binaries. 
\textit{Right panel}: mass-dependent binary fraction ($f_{\rm bin}$) in our realisation (blue points) compared to observational constraints from \citet{moe17}.}
    \label{fig:moe}
\end{figure*}

First, we generate a population of $N$ single stars and binaries and the total number of stars is $(1+f_{\rm bin})N$, where $f_{\rm{bin}}$ is the binary fraction. Stellar masses are sampled from a \citet{kroupa2001} initial mass function (IMF)  in the range 0.08-150 $M_\odot$. Masses of primary stars in binaries are drawn from the IMF as single stars, while secondary components are paired based on the mass ratio distribution.  Stellar and binary properties are assigned according to the empirical prescriptions of \citet{moe17}, which provide mass-dependent distributions for the binary fraction, mass ratio, period, and eccentricity (see Fig.~\ref{fig:moe}). Second, we compute the total stellar mass $M_{\rm SC}$ by summing the masses of all stars, including both components of binaries.
Third, we sample $N$ phase space coordinates for the single stars and for the center-of-mass of the binaries for each cluster using a custom Python pipeline 
based on the \texttt{limepy} library \citep{gieles2015}, which samples positions and velocities from equilibrium dynamical models with a specified central potential. We adopt a King model \citep{king66} with dimensionless central potential $W_{0} = 7$ and total mass $M_{\rm SC}$. 

For binaries, we additionally compute the relative positions and velocities of the two components based on their sampled orbital parameters. The orbital phase is specified by drawing the mean anomaly uniformly from $[0, 2\pi]$, the inclination by sampling $\cos{i}$ uniformly in $[-1, 1]$, and the longitude of the ascending node uniformly in $[0, 2\pi]$.

We consider two different metallicities: $Z = 0.0002$ (metal poor) and $Z = 0.02$ (metal rich) where $Z$ indicates the mass fraction of metals (all elements heavier than helium).
In total we ran 3600 $N$-body  simulations.
Since each SC hosts a population of binaries built up during cluster initialization, we distinguish two classes of binaries: i) \textit{original binaries}, whose components have been bound since the cluster's formation (see Fig. \ref{fig:moe}), and ii) \textit{exchanged binaries}, where the components become bound through dynamical encounters during the cluster's evolution.

\subsection{$N$-body simulations}
\label{sec:nbody}

To follow the dynamical evolution and the interactions between stars and COs in the generated SCs, we perform a suite of $N$-body simulations with the state-of-the-art code \pt \citep{Wang2020b}.
\pt combines a fourth-order Hermite integrator for short-range interactions with the Particle-tree  method for long-range interactions 
\citep{oshino2011} to efficiently handle both close encounters and large-scale dynamics. 
\pt is designed to work across multiple nodes on supercomputer clusters taking advantage of a hybrid parallelization: MPI, OpenMP CUDA, SIMD instructions (AVX, AVX2, AVX-512), and GPU acceleration are exploited to reduce the computational effort through \fdps platform \citep{Iwasawa2016,Iwasawa2020}. An accurate regularization algorithm (\sdar, \citealt{Wang2020a}) is implemented in the code to handle close encounters, binaries, and hierarchical systems. 

However, the current version of \pt does not include any prescriptions for micro-TDEs. To address this, we have developed a customized version of the code that incorporates them. 
In our \pt version, when a star falls within $r_{\rm t}$ of a CO (see Eq. \ref{eq:cond}), \pt removes the star from the simulation and accretes 10\% of its mass onto the CO. Additionally, we have improved the overall treatment of object collisions, resolving an issue that could previously cause spurious immediate collisions between distant objects (see Appendix \ref{app:petar} for details).

\subsubsection{Stellar evolution and binary processes} \label{sec:bse}
In its default version, \pt accounts for stellar evolution and binary processes through a coupling with the population synthesis code \bse \citep{hurley00,hurley02} in the updated version described in \citet[Section~2]{banerjee20}. In particular, the updates incorporate new semi-empirical stellar wind prescriptions for massive stars following \cite{B10}, supernova (SN) and remnant formation models by \cite{fryer12}, pair-instability and pulsation pair-instability SN models as in \cite{B16}.

In our simulation we use the delayed SN model by \cite{fryer12} in which the explosion is driven by neutrinos  on a time scale of hundreds of milliseconds after the formation of the proto-NSs. These models imply a continuous distribution of remnant masses, with no mass gap between NSs and BHs consistent with the recent detection of massive NSs or low-mass BHs in the mass range 2.5-5 M$_\odot$ \citep[see e.g.][]{Wyrzykowski2020,GW190814,Ray2025}. In our models the minimum mass allowed for BH is thus 3 \Ms.

\subsubsection{External tidal field and  initialisation} \label{sec:efield}

In order to evolve the SCs within a realistic Galactic tidal
field, we include a MW-like external potential based on the
model \texttt{MWPotential2014} in \texttt{GALPY} introduced in \cite{bovy}.
The dynamical model includes a inner spheroidal bulge-like component with total mass $M_\mathrm{bulge}=5 \times 10^9 \ M_\odot$, a Myamoto-Nagai  stellar disc \citep{MNdisc} with mass $M_\mathrm{disc}=6.8 \times 10^{10} \ M_\odot$, and a Navarro-Frank-White  dark matter halo \citep{NFW} with virial mass  $M_{\rm halo}=8 \times 10^{12} \ M_\odot$.

 Each cluster is initially placed on a circular orbit ($220\,\mathrm{km\,s}^{-1}$) at the Solar radius ($R = 8\,\mathrm{kpc}$).
 The clusters are then evolved for $\approx 1.5$ Gyr, ensuring that all the SCs experience at least one relaxation time during the $N$-body integration.

\section{RESULTS}
\label{sec:res}
\subsection{Micro-TDE classification}
 \label{sec:channels}
We identify three main channels involving COs and stars that satisfy the TDE condition (Eq.~\ref{eq:cond}) and lead to micro-TDEs.
 \begin{itemize}
     \item "Singles": the single CO and the single star are not previously bound and approach each other on a hyperbolic/parabolic orbit that brings the star at a pericenter distance which is within the CO's $r_{\rm t}$\footnote{This scenario is analogous to classical TDEs involving SMBHs.}.
     \item "Binaries": this channel includes cases in which a star and a CO are bound together and Eq.~\ref{eq:cond} is satisfied because of i) binary hardening following distant dynamical interactions; or
     ii) the expansion of the star in the binary (e.g. stellar radius expansion etc.); or iii) the CO progenitor undergoes a SN explosion \citep{perets2016, michaely2016,hirai2022,Tsuna2025}. If the newly formed CO receives a sufficiently strong natal kick, the binary orbit may be altered in such a way that the pericenter of the surviving star falls within $r_{\rm{t}}$, leading to a micro-TDE. These events are grouped under the name "SN-kick" sub-channel.
     
     \item "Multiples": this channel involves micro-TDE induced during multiple dynamical interactions \citep{perets2016} as i) \textit{hierarchical} triples (or quadruples), where an inner binary is perturbed by one (or more) outer component(s); and  ii) \textit{democratic/resonant} few-body encounters, in which the star is disrupted during a non-secular $N$-body interaction involving three or more components.     
     \end{itemize}

BH-star binaries that undergo a micro-TDE can either form dynamically during the cluster evolution or originate as bound systems at birth. Accordingly, when discussing the binaries and multiples channels, we further distinguish whether the BH-star pair system involved in the micro-TDE is an \ex or an \og binary.

\subsection{The "singles" channel}
\label{sec:hyper}

\begin{figure}
 	  \centering
    \includegraphics[width=0.45\textwidth]{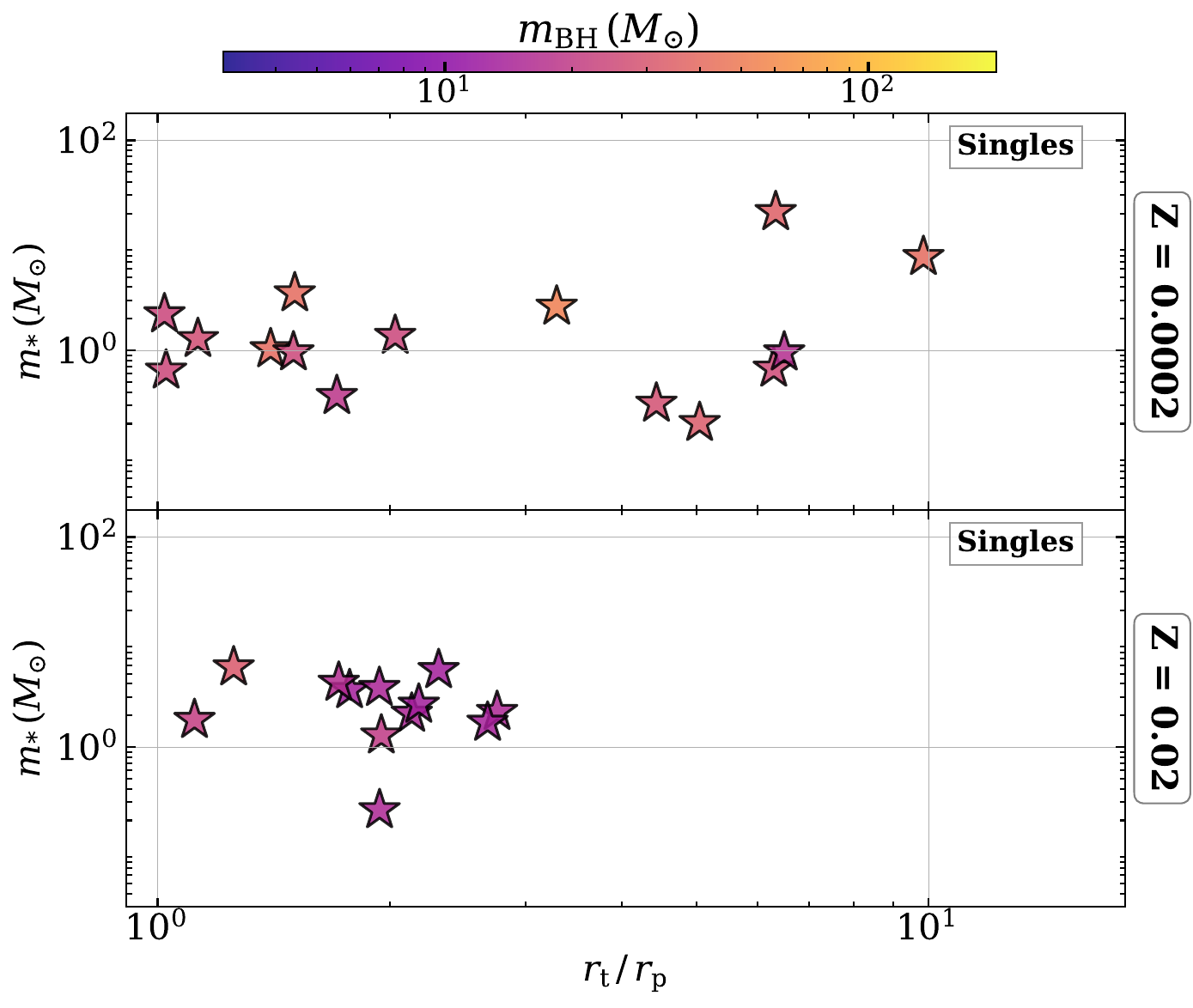}
   \caption{Distribution of \be ($r_{\rm{t}}/r_{\rm{p}}$) and star masses ($m_{*}$ in \Ms) of micro-TDEs occurring during single encounters for $Z = 0.0002$ (\textit{top panel}) and $Z = 0.02$ (\textit{bottom panel}). The colormap shows the BH masses ($m_{\rm{BH}}$ in \Ms).}
    \label{fig:hyp}
\end{figure}

Considering both metallicities and all formation channels, only about $\approx 3$ \% of micro-TDEs originate from BH-star single encounters. 
In Figure~\ref{fig:hyp}, we show the distribution of BH and star masses and \textit{penetration factor}, defined as $\beta = r_{\rm t} / r_{\rm p}$, of single encounters for metal-poor (\textit{top panel}) and metal-rich (\textit{bottom panel}) YSCs.

At $Z=0.0002$, the BH masses ($m_{\mathrm{BH}}$) range from $\approx17$ to $\approx52$~\Ms, while the stellar companion masses span $\approx0.2$ to 21~\Ms.
 At solar metallicity, because of strong stellar winds \citep{vink01}, the BH masses involved are typically lower, ranging in this sample from $\approx13$ to $\approx31$~\Ms while the disrupted stars have masses between $\approx13.3$ and $\approx13.8$~\Ms. 
 A notable difference between the two metallicities appears in the distribution of \textit{penetration factors}. In metal-rich YSCs, the encounters typically have $r_{\rm{t}}/r_{\rm{p}} \approx 2-3$, indicating that the interactions are generally less deeply penetrating compared to metal-poor environments ($r_{\rm{t}}/r_{\rm{p}}$ up to $\approx 10$). All interactions are in the gravitational focusing regime, such that the pericenter of a hyperbolic or parabolic encounter is inversely proportional to the total mass of the interacting objects. Because the mass distribution of the disrupted stars is not strongly dependent on metallicity, the range of \be is primarily determined by the mass of the BHs. As shown in the bottom panel of Fig. \ref{fig:hyp}, at \Zs, most events occur with BH masses clustered around \( \sim 15 \, {\rm M}_\odot \), while at \( Z = 0.0002 \), the range of BH masses is broader and including more massive BHs, leading to a wider spread in \be, ranging from \( \sim 1 \) to \( \sim 10 \).
 
 Most stars involved in single encounters are MS stars with few exceptions (5\% giants) at both metallicities.
 Roughly 40\% single encounters take place within the core radius of the parent SCs, where high density favour these kind of interactions.

\begin{figure}
 	  \centering
\includegraphics[width=0.45\textwidth]{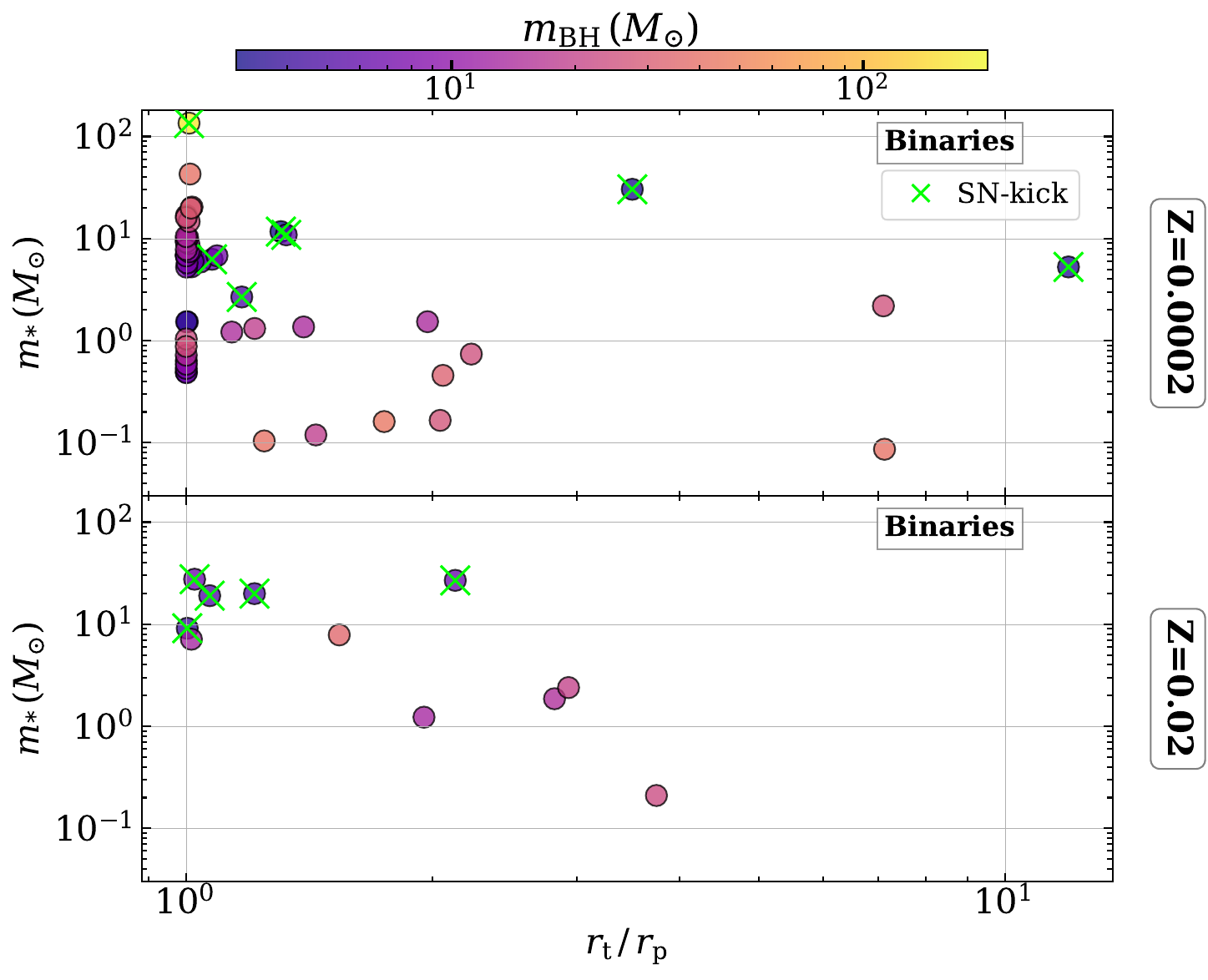}
\caption{Same as Fig. \ref{fig:hyp} but for the channel "binaries". The green cross indicate SN-trigger events. }
    \label{fig:binaries}
\end{figure}

\subsection{The "binaries" channel}
\label{sec:bin}

Micro-TDEs originating from the "binaries" channel are rare, accounting for $\approx7$\% of the total sample, and occur predominantly ($\approx 82$\%) in metal-poor environments (see also Sect.~\ref{sec:eta}).

At $Z = 0.0002$, these events involve BH masses ranging from $\approx 3$ to $\approx 170$~\Ms{} and stellar masses from $\approx 0.1$ to $\approx 134$~\Ms. 
The disrupted stars are mainly MS stars ($\approx 35$\%), followed by giants and naked helium stars\footnote{We define naked helium stars as stars that have lost their hydrogen envelope either through strong stellar winds (typical in metal-rich environments) or binary interactions. See \citet{hurley00,hurley02} for further details.}, each contributing $\approx 50$\%.

In metal-rich environments ($Z = 0.02$), BHs involved in this channel are typically lighter, with masses between $\approx 4$ and $\approx 35$~\Ms, and stellar masses between $\approx 0.2$ and $\approx 27$~\Ms. 
Here, the disrupted stars are exclusively MS stars.

Among both metallicities, approximately $70$\% of the events occur with a \be close to $1$. These are cases in which the TDE condition (Eq.~\ref{eq:cond}) is triggered by the expansion of the companion star’s radius.
Overall, $\approx 70$\% of the micro–TDEs in the binary channel arise from the evolution of \og binaries (see Sect.~\ref{sec:ic}), driven primarily by stellar processes such as common envelope (CE) evolution and mass transfer (MT), rather than dynamical interactions.

This evolutionary origin is also reflected in the eccentricity distribution: for the binary channel, the distribution peaks around $e \approx 0.1$, with $\approx 42$\% of systems falling in this range, consistent with the circularization effects expected from CE and MT phases.

A subset of micro-TDEs in the binary channel is triggered by the SN explosion of the BH progenitor \citep{perets2016,Tsuna2025}. These events are rare, accounting for $\approx 0.8$\% of the total micro-TDEs at $Z = 0.0002$ and $\approx 0.5$\% at \Zs. However, within the binary channel at Solar metallicity, nearly half of the events are associated with SN natal kicks (see Fig.~\ref{fig:binaries}, green crosses).

In metal-rich environments, the larger stellar radii lead to early stellar mergers that prevent the formation of BH-star systems in close orbits \citep[see e.g.][]{iorio2023}. The surviving BH-star binaries reside in wide systems where a fortunate post-SN kick configuration can place the stars on an eccentric orbit, thereby triggering the micro-TDE.

\subsection{The "multiples" channel}
\label{sec:mult}

\begin{figure}
 	  \centering
\includegraphics[width=0.45\textwidth]{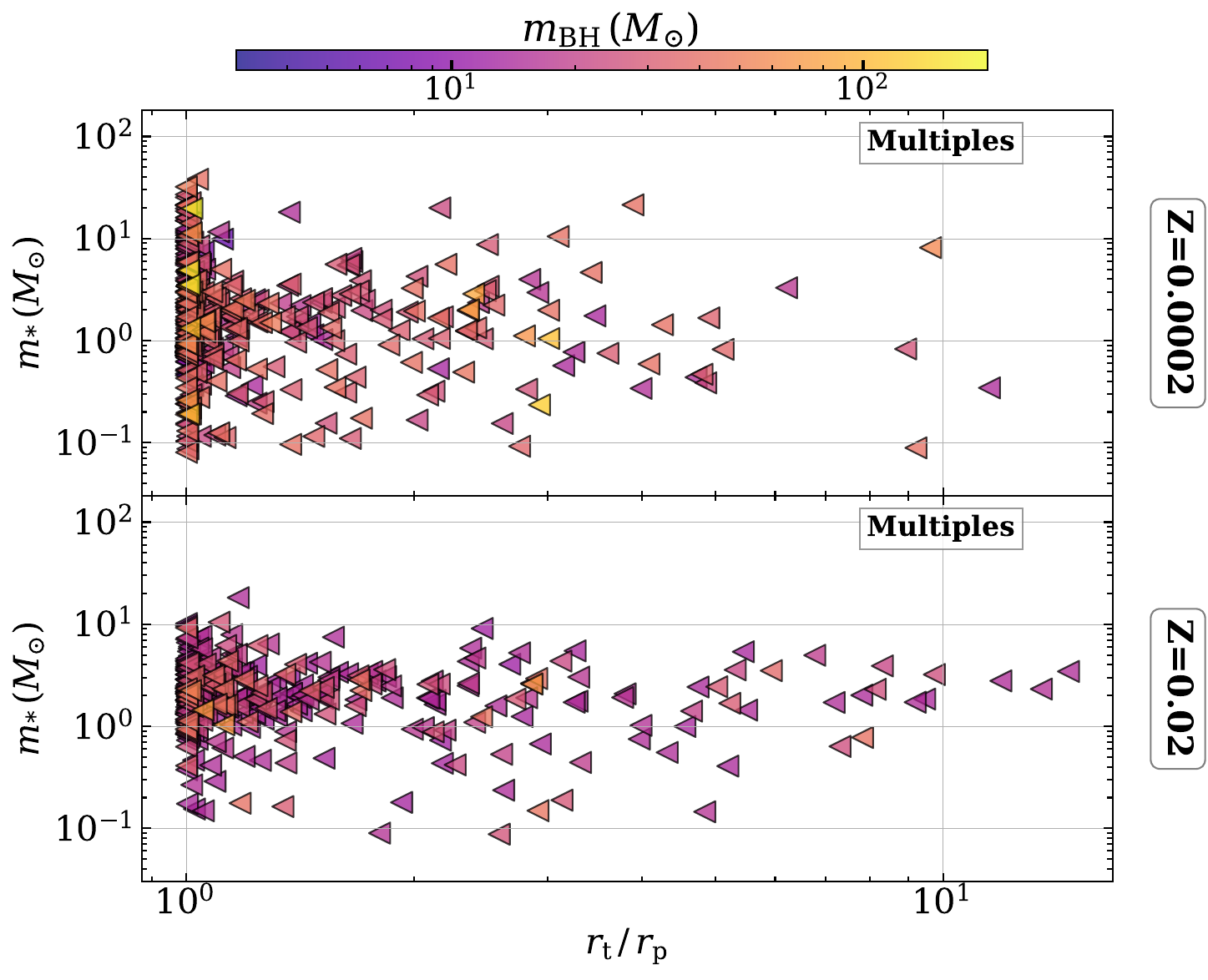}
   \caption{Same as Fig. \ref{fig:binaries} but for the channel "multiples".}
    \label{fig:multipl}
\end{figure}

Most (\(\approx 90\%\)) micro-TDEs occur during multiple encounters within the evolution of YSCs (Fig.~\ref{fig:multipl} ): 85\% in metal-poor and 90\% in metal-rich clusters. Notably, 91\% of these multiple interactions involve \ex BH-star binaries, highlighting the importance of dynamics in shaping these events.

At \(Z = 0.0002\), the stellar masses range from \(\approx 0.08\)~M\(_\odot\) to \(\approx 38\)~M\(_\odot\), and BH masses range from 3~M\(_\odot\) up to \(\approx 138\)~M\(_\odot\). Nearly all events (\(> 99\%\)) occur on nearly parabolic orbits (\(e \approx 0.99\)), with the majority involving MS stars (83\%), followed by giants (8\%), naked helium stars (7\%), and WD (0.5\%).\\
Similarly, at \(Z = 0.02\), micro-TDEs predominantly involve MS stars (98\%), followed by a small fraction of giants (0.9\%) and WD (1.1\%). 
The stellar mass ranges from a minimum of \(\approx 0.09\)~M\(_\odot\) to a maximum of \(\approx 18\)~M\(_\odot\), while the BH, as expected, are much lighter compared to metal-poor YSCs, with masses ranging from \(\approx 11\)~M\(_\odot\) to \(\approx 75\)~M\(_\odot\).

As shown by comparing Fig.~\ref{fig:binaries} and Fig.~\ref{fig:multipl}, the latter spans a wider range of $\beta$, indicating that as a consequence of multiple dynamical interactions, stars and BHs in YSCs can approach each other more closely, resulting in highly penetrating encounters within the BH tidal radius.

Within the multiples, we categorize hierarchical triples and quadruples as flagged by the \pt\ code. We classify their configurations based on the nature of the outer object (first term) and the components of the inner binary (enclosed in square brackets).

At $Z=0.0002$, the most common triple configurations leading to stellar disruption are those involving COs in the inner binary. The dominant channel is BH–[BH–Star], accounting for 60\% of the events, followed by Star–[BH–Star] (30\%). Less frequent are BH–[Star–Star] (4\%) and the rare case of Star–[BH–BH] (1\%).

At $Z = 0.02$, most disrupted stars are found in triples of the form Star–[BH–Star] (57\%), followed by BH–[Star–BH] (36\%) and BH–[Star–Star] (7\%).

In the case of quadruples, a direct classification into inner and outer components is less straightforward. We therefore characterize them by counting the number of BHs present in the systems. At $Z = 0.0002$ ($Z = 0.02$), the majority of disrupted systems contain one BH (56.00\%, 82\%), followed by systems with two BHs (23\%, 17\%). Systems hosting three BHs are more frequent at low metallicity (21\%) but are rare at solar metallicity (1.2\%).

We note that we do not find any triple or quadruple hosting both BHs and NSs simultaneously.

\subsection{Micro-TDE production efficiency} \label{sec:eta}

We assume that the number of  micro-TDEs in a stellar cluster with mass $M_\mathrm{SC}$ follows a Poisson distribution with mean $\lambda_\mathrm{Poisson}=\eta M_\mathrm{SC}$, where $\eta$ is the micro-TDE production efficiency.

Consequently, if $N_\mathrm{micro-TDE,tot}$ are produced across clusters with a combined stellar mass $M_\mathrm{SC,tot}$, the posterior distribution for $\eta$ follows a Gamma distribution:
\begin{equation}
P(\eta|N_\mathrm{micro-TDE},M_\mathrm{SC,tot}) \sim \mathrm{Gamma}(\alpha + N_\mathrm{micro-TDE,tot}, \gamma + M_\mathrm{SC,tot} ),
\label{eq:posteta}
\end{equation}
where we assume a prior $P(\eta) \sim \mathrm{Gamma}(\alpha, \gamma) \propto \eta^{\alpha -1} e^{-\gamma \eta}$ \citep{Gelman2014}. 

In the limit of large $N_\mathrm{micro-TDE,tot}$, the posterior distribution converges to a Gaussian with mean $N_\mathrm{micro-TDE,tot} / M_{\mathrm{SC,tot}}$ and standard deviation $\sqrt{N_\mathrm{micro-TDE,tot}} / M_{\mathrm{SC,tot}}$. However, for small sample sizes, Eq.~\ref{eq:posteta} provides a more robust and statistically rigorous estimate of $\eta$ and its associated uncertainty.

We use Eq.~\ref{eq:posteta} to estimate the global value of $\eta$ from our simulations, and to explore its variation as a function of cluster mass, density, metallicity, and micro-TDE formation channel.
For the prior, we adopt $\alpha = 1$ and 
$\gamma = 0.1 \ \mathrm{M}_\odot$,  corresponding to a nearly flat, uninformative distribution over the range $\eta < 1  \ \mathrm{M}_\odot^{-1}$ $\left( P(\eta) \propto e^{-\eta/10 \,\mathrm{M}^{-1}_\odot }\simeq1 \right)$.

\begin{figure}
 	  \centering
\includegraphics[width=0.45\textwidth]{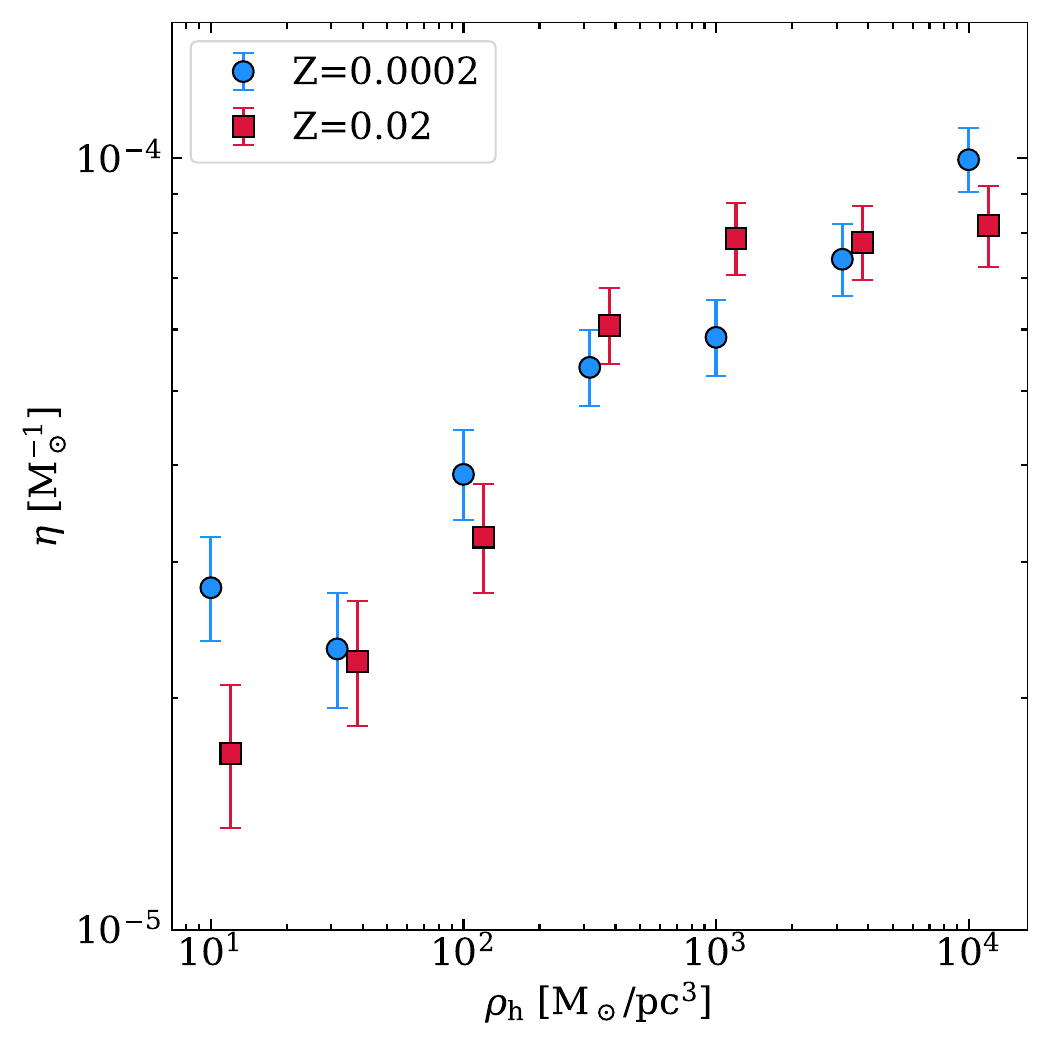}
   \caption{Micro-TDE production efficiencies, $\eta$, as a function of initial cluster density $\rho_{\rm h}$ for $Z = 0.0002$ (blue) and $Z = 0.02$ (red). The seven points correspond to cluster densities sampled from the cluster initial conditions (see Figure \ref{fig:kru}). The red points have been slightly shifted to improve visual clarity. Each point and its error bar represent the median and the 68\% credible interval from the posterior distribution defined in Equation \ref{eq:posteta}. }
    \label{fig:eff_rho}
\end{figure}

Considering the full simulation dataset, the total production efficiency is $\eta \approx 5 \times 10^{-5} \,\mathrm{M}_\odot^{-1}$. 
Figure~\ref{fig:eff_rho} shows an increasing trend of $\eta$ as a function of the cluster density. 
The most extreme values are observed in low-density clusters ($\approx 2 \times 10^{-5} \ \mathrm{M}_\odot^{-1}$) and high-density clusters ($ \approx 9 \times 10^{-5} \ \mathrm{M}_\odot^{-1}$). The dependence on cluster mass is less pronounced, with the highest efficiency ($\approx 6 \times 10^{-5}~\mathrm{M}_\odot^{-1}$) observed in intermediate-mass clusters ($\approx 5 \times 10^{4}~\mathrm{M}_\odot$), while the efficiency decreases to $\lesssim 3 \times 10^{-5}~\mathrm{M}_\odot^{-1}$ in the most massive clusters.
 The fact that $\eta$ peaks at intermediate masses is the result of the scaling of the interaction rate with $M_{\rm SC}$, $\rho_{\rm h}$ and two-body relaxation: in the gravitational focussing regime, the interaction rate per unit mass is proportional to $\rho_{\rm h}/\sigma_{\rm rms}$, where $\sigma_{\rm rms}$ is the velocity dispersion. In virial equilibrium $\sigma_{\rm rms}\propto \rho_{\rm h}^{1/6}M_{\rm SC}^{1/3}$ such that in the absence of dynamical evolution $\eta$ is expected to scale with cluster properties as $\eta\propto\rho_{\rm h}^{5/6}M_{\rm SC}^{-1/3}$, that is, the lowest $\eta$ for the most mass clusters (at a given $\rho_{\rm h}$). Two-body relaxation leads to an expansion of low-mass clusters, such that  $\rho_{\rm h}\propto M^2$ \citep[e.g.,][]{gieles2011}, such that after some relaxation we have at low masses: $\eta\propto M_{\rm SC}^{4/3}$. Hence, the lowest mass clusters expand as the result of relaxation leading to a reduction of $\eta$, while the most massive clusters have a low $\eta$ because of their high velocity dispersion.

\begin{figure}
 	  \centering
    \includegraphics[width=0.45\textwidth]{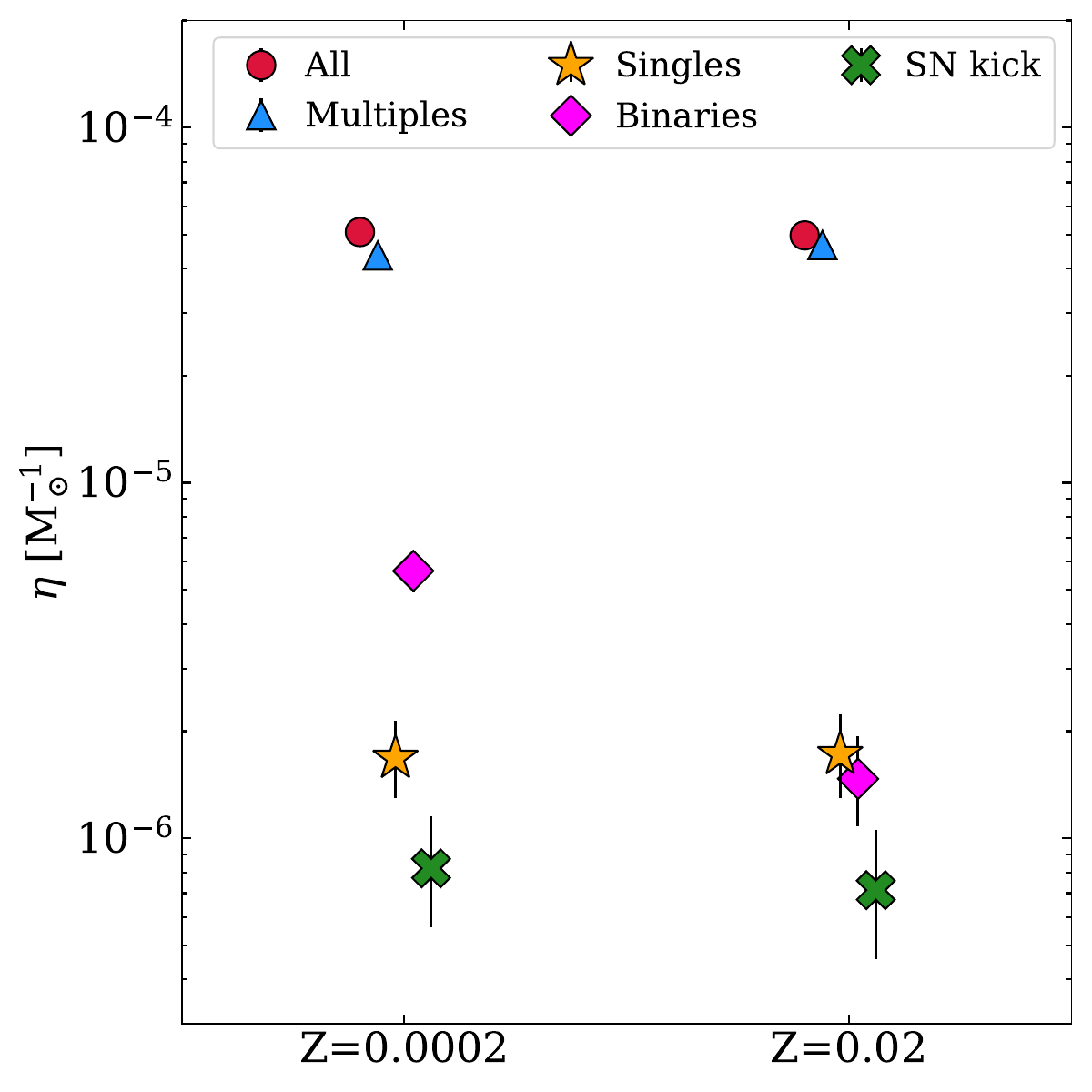}
   \caption{Micro-TDE production efficiency for the two simulated metallicities. Circles indicate the efficiency considering the entire sample of events, while the other symbols represent contributions from different channels: diamonds for single interactions, stars for binaries, triangles for multiples, and crosses for micro-TDEs triggered by SN kicks (see Section~\ref{sec:channels}). Each point and its error bar represent the median and the 68\% credible interval from the posterior distribution defined in Eq. \ref{eq:posteta}. }
    \label{fig:eff_ch}
\end{figure}

Figure~\ref{fig:eff_ch} shows that $\eta$ is independent of metallicity across all the micro-TDE channels analyzed, except for the binary channel, which drops from $6 \times 10^{-6} \ \mathrm{M}_\odot^{-1}$ at low metallicity to $2 \times 10^{-6} \ \mathrm{M}_\odot^{-1}$ at high metallicity (see also Fig.~\ref{fig:binaries}).
In fact, metal-rich stars have larger stellar radii 
\citep[see e.g.][]{costa2025}, implying a higher likelihood of premature stellar mergers before the formation of the first compact remnant, thus preventing the formation of BH-star systems, hence lowering the rate of micro-TDE binaries in metal rich environment.
Micro-TDEs induced by SN kicks account for the lowest contribution, with $\eta \approx 7 \times 10^{-7} \ \mathrm{M}_\odot^{-1}$.
The production efficiency of single encounters is subdominant ($\approx 1.7 \times 10^{-6}\,\mathrm{M}_\odot^{-1}$) in both environments. 

The dominant contribution to the micro-TDE population at both metallicities comes from dynamically formed systems, particularly those involving higher-order multiples, with a production efficiency of $\eta \approx 5 \times 10^{-5} \ \mathrm{M}_\odot^{-1}$. Among the three formation channels, the multiple channel account for 86\% of all micro-TDEs at $Z = 0.0002$ and 94\% at $Z = 0.02$ (Fig.~\ref{fig:multipl}).
 The large number of dynamical interactions occurring in dense YSCs favors the formation of few-body systems (triples or quadruples) and promotes interactions with \og binaries, thereby enhancing encounters between stars and BHs and thus micro-TDEs in multiples.
 
 Additionally, the initial fraction of \og binaries (Sec. \ref{sec:ic}) is another crucial ingredient because stellar binaries are known to increase the rates of dynamical collisions and TDEs (e.g., \citealt{fregeau2007}). 
\textit{Original} binaries segregate toward the central region of SCs where high density enhance the interactions, collisions and three-body/multi-body encounters which boost the formation of \ex binaries and the reshuffling of the \og binary components, creating new multiple systems and/or hardening the \og binaries enough to satisfy the TDE condition (Eq. \ref{eq:cond}).

The low dependence on metallicity and high dependence on cluster density highlight  that the overall micro-TDE production is highly sensitive to the cluster's dynamical environment, with higher-order multiple and few-body interactions playing a central role in driving stellar disruptions.

Cluster density, which has the strongest impact on production efficiency, is also one of the most poorly constrained initial parameters for YSCs \citep{krumholz2019}. In our simulations, we explore a wide range of densities spanning three orders of magnitude, and we use the production efficiencies obtained at the lowest ($\rho_{\rm h} = 1 \ \mathrm{M}_{\odot}/\mathrm{pc}^3$) and highest ($\rho_{\rm h} = 10^4 \ \mathrm{M}_{\odot}/\mathrm{pc}^3$) density values to bracket the systematic uncertainties. In these two extreme cases, the results reflect the efficiency of micro-TDE production under the assumptions that all clusters are either born with a diffuse structure, making them more susceptible to disruption, or are significantly more concentrated, and thus more resilient to disruption and characterized by enhanced dynamical activity.

A summary of $\eta$ posteriors for different production channels and for different cluster subset can be found in  Table \ref{tab:eta} in the appendix.

\subsection{Micro-TDE Event Rates} 
\label{sec:rates}

Using the estimated $\eta$ (Section \ref{sec:eta}), we derived the comoving event rate density, $R(z)$, which quantifies the number of micro-TDE per unit comoving volume and per unit time as a function of redshift.
Considering the modest variation of the production efficiency with  metallicity and cluster mass Figure \ref{fig:eff_ch} and Tab.~\ref{tab:eta}), we simplified the estimate of comoving rates by neglecting the metallicity and cluster mass dependence. 
In addition, we assumed a null delay time, i.e. we assume that all the micro-TDE events  are triggered at the moment of the cluster formation. 
This assumption is justified by the fact that in all simulated clusters, 50\% of the TDEs are produced within 300 Myr, and 80\% within 800 Myr (see also Sect. \ref{sec:time}), so we expect only a moderate variation in the redshift evolution of the rate with respect to the assumed star formation rate density (SFRD), $\rho_\mathrm{SFRD}(z)$.
With these approximations, the rate density as a function of redshift is 
\begin{equation}
R(z)  = \eta f_\mathrm{SF,SC} \rho_\mathrm{SFRD}(z),
\label{eq:rate}
\end{equation}
where, $f_\mathrm{SF,SC}$ is the fraction of total star formation occurring in YSCs.
We treated $\eta$ as a stochastic variable, following the distribution given in Equation~\ref{eq:posteta}. 
For $f_\mathrm{SF,SC}$ a commonly adopted value in the literature is $0.8$ (see, e.g., \citealt{Lada2003,kremer2021}), but this parameter remains poorly constrained. We therefore modeled it as an additional stochastic variable, assuming a Beta distribution with parameters $\alpha_B = 4.17$ and $\beta_B = 1.60$ \citep{Gelman2014}. This distribution peaks at $f_\mathrm{SF,SC} = 0.8$, has a standard deviation of 0.15, and vanishes as $f_\mathrm{SF,SC} \to 1$.
For the star formation rate density (SFRD), we adopted the redshift-dependent model from \citet{Madau2017}:
\begin{equation}
\rho_\mathrm{SFRD}(z) = 10^7\,{\rm M}_\odot\,{\rm yr}^{-1} \,{\rm Gpc}^{-3} \frac{(1+z)^{2.6}}{1 + \left(\frac{1+z}{3.2}\right)^{6.2}}. 
\label{eq:sfrd}
\end{equation}

\begin{figure}
 	  \centering    \includegraphics[width=0.45\textwidth]{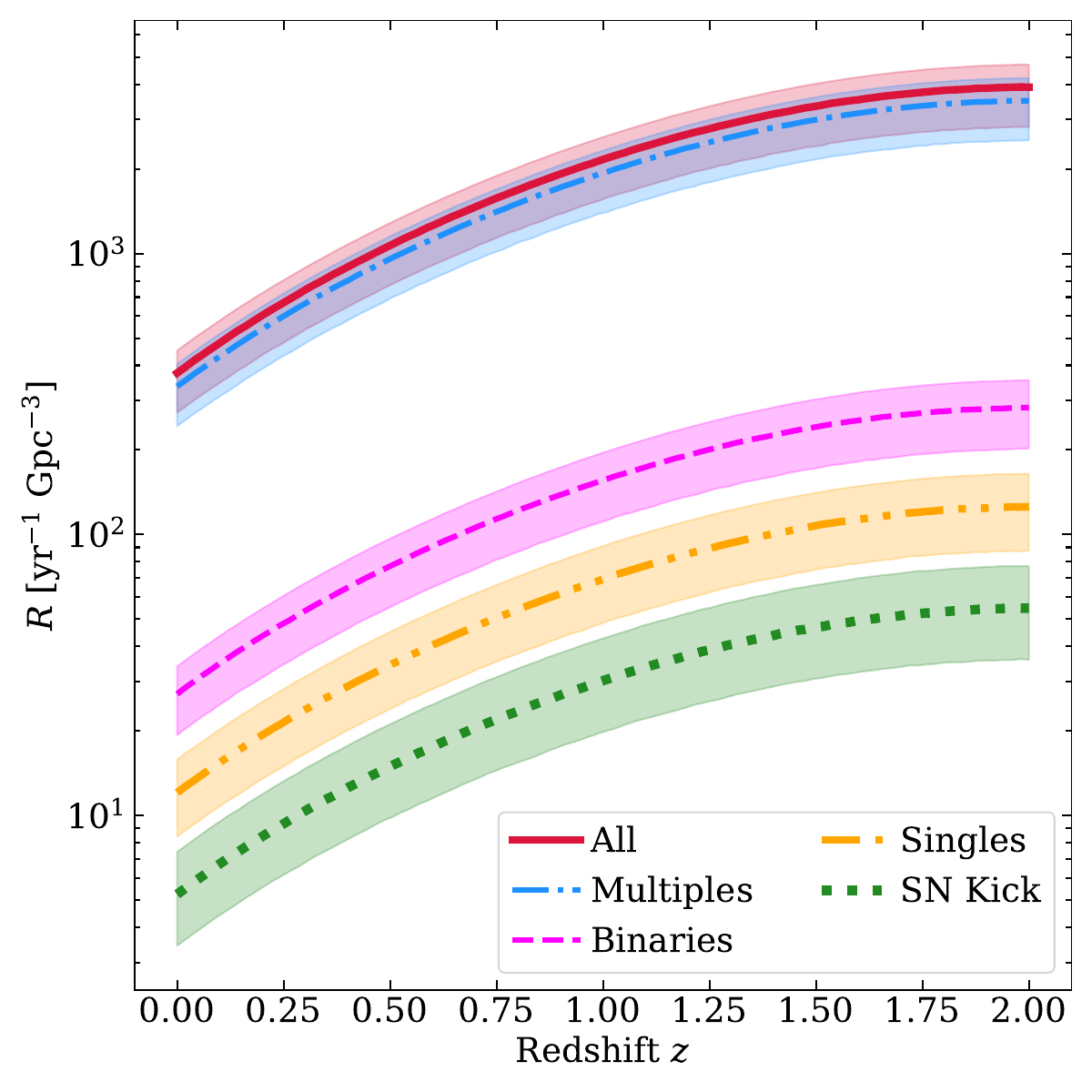}
   \caption{Micro-TDE volumetric rates as a function of redshift (Eq.~\ref{eq:rate}). Solid lines show the median rates, while shaded bands indicate the 68\% credible intervals obtained by sampling the posterior distributions of the production efficiency (Eq~\ref{eq:posteta}) and the fraction of star formation occurring in clusters (see Sect.~\ref{sec:rates}). The red solid line corresponds to the total rate, while colored lines represent the contributions from different formation channels: singles (orange, dash-dot-dot), binaries (magenta, dashed), multiples (blue, dash-dot), and SN-kicks (green, dotted), as detailed in Sec.~\ref{sec:channels}.}
    \label{fig:rate}
\end{figure}

At each redshift, we sampled the posterior distribution of the event rate density by drawing $10^4$ values for both $\eta$ and $f_\mathrm{SF,SC}$. Figure~\ref{fig:rate} shows the resulting micro-TDE event rate density as a function of redshift, for the total population as well as for the different production channels (Section~\ref{sec:channels}). Following the trend of the cosmic star formation rate density, $R(z)$ increases by a factor of approximately 5 from $z = 0$ to $z = 1$, and by a factor of approximately 10 from $z = 0$ to $z = 2$.

By integrating the event rate density over the comoving volume, $V_{\rm c}$, we obtained the cumulative event rate up to redshift $z$:
\begin{equation}
\Gamma(z) = \int^z_0 R(z)  \frac{d V_{\rm c}}{d z} \frac{1}{1+z}  dz = \eta f_\mathrm{SF,SC} \int^z_0 \rho_\mathrm{SFRD}(z) \frac{d V_{\rm c}}{d z} \frac{1}{1+z}  dz.
\label{eq:number}
\end{equation}

To evaluate the posterior distribution of $\Gamma(z)$, we estimated the event rate density by drawing $10^4$ samples for both $\eta$ and $f_\mathrm{SF,SC}$ (Equation~\ref{eq:rate}). As our fiducial model, we adopted the distribution derived from the full sample of simulated YSCs. Systematic uncertainties were bracketed by considering the extreme values obtained from low-density ($10 \ \mathrm{M}_\odot\, \mathrm{pc}^{-3}$) and high-density ($10^4 \ \mathrm{M}_\odot \,\mathrm{pc}^{-3}$) cluster populations (see Section~\ref{sec:eta} and Table~\ref{tab:eta}).
The integral over comoving volume in Equation~\ref{eq:number} was evaluated numerically, assuming the Planck18 cosmology \citep{Planck18}.\footnote{We estimated the integral using the quadrature method \texttt{quad} from the \texttt{Python} module \texttt{scipy} \citep{scipy}, and we computed the differential comoving volume using the function \texttt{differential\_comoving\_volume} from the \texttt{astropy} module \citep{astropy}. }

\begin{figure}
 	  \centering
\includegraphics[width=0.45\textwidth]{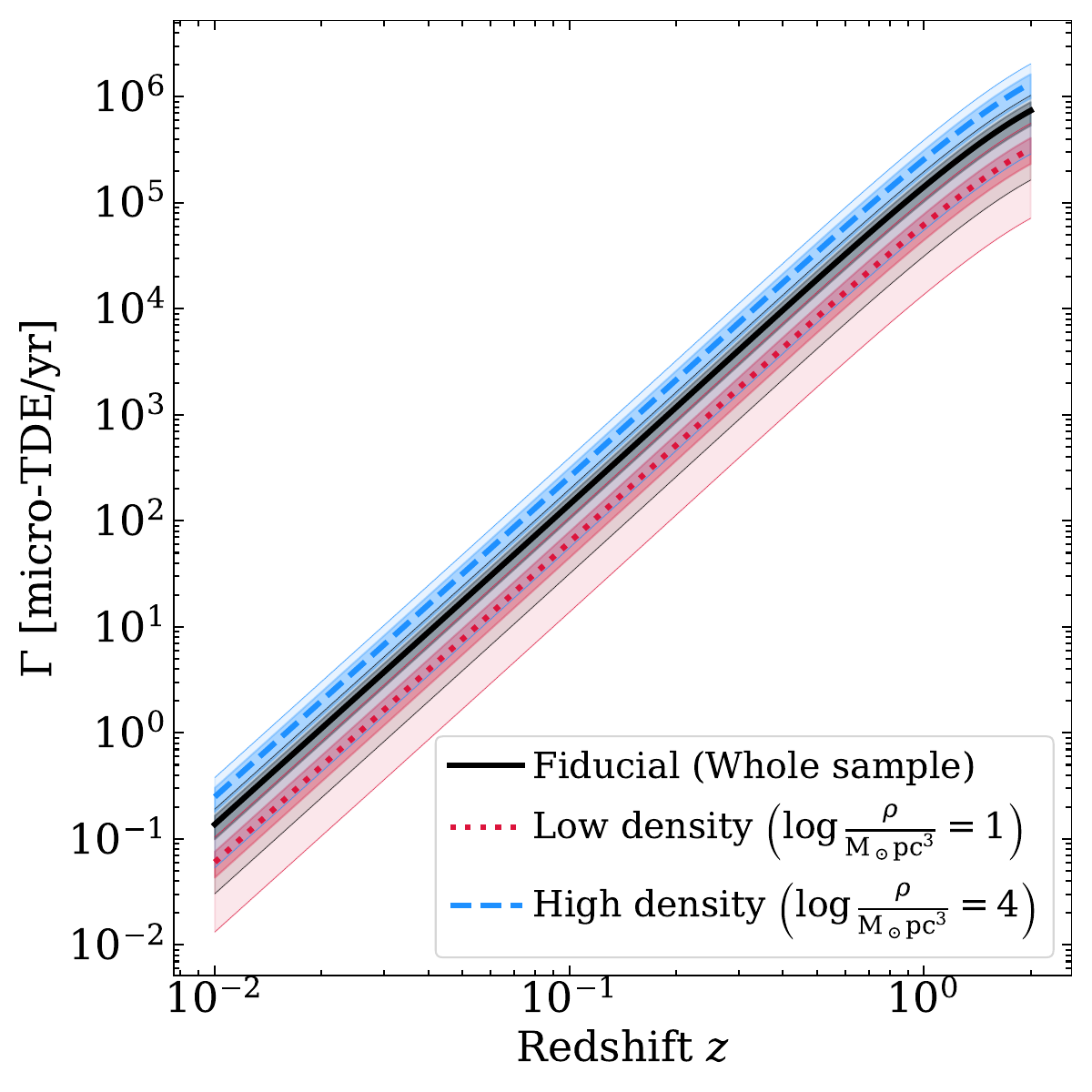}
   \caption{Cumulative micro-TDE rate as a function of the redshift (Equation \ref{eq:number}). The colors indicate the fiducial (black), pessimistic (low density, red) and optimistic (high density, blue) model for the micro-TDE production efficiency (see Table \ref{tab:eta}). The lines indicate the median, while the bands the 68\% (dark) and 99.7\% (light) credible interval at a given redshift z.}
    \label{fig:ntde}
\end{figure}

Figure~\ref{fig:ntde} shows the expected cumulative number of micro-TDE events per year for the fiducial model based on the full cluster sample as well as for the two limiting cases corresponding to low- and high-density clusters. The number of expected events per year increases from just a few within the local Universe ($z < 0.05$, approximately 200 Mpc) to up to $10^5$ at $z = 1$.
The dominant sources of uncertainty are the poorly constrained initial density distribution of stellar clusters (see Section~\ref{sec:eta}) and the assumed fraction of star formation occurring in YSCs. Within the explored range of cluster densities ($10$–$10^4 \,\mathrm{M}_\odot \,\mathrm{pc}^{-3}$; see Section~\ref{sec:ic}) and adopting the sampled distribution for $f_\mathrm{SF,SC}$, the $3\sigma$ uncertainty spans approximately 1.7 dex, corresponding to a maximum-to-minimum ratio of $\sim$50.

\subsection{Micro-TDEs on NSs}
\label{sec:NS} 
\begin{figure}
 	  \centering
\includegraphics[width=0.45\textwidth]{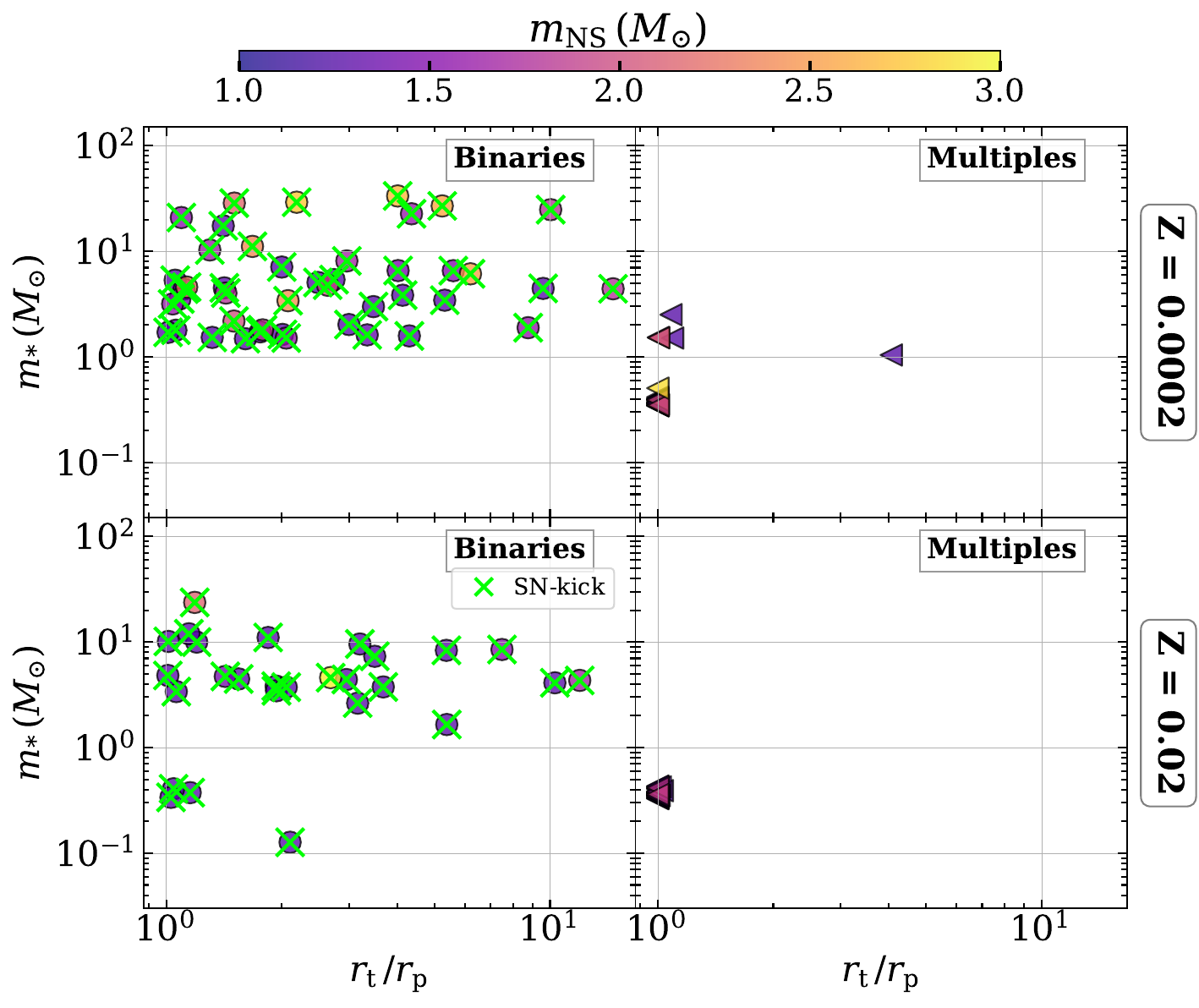}
   \caption{Same as Fig. \ref{fig:binaries} but for micro-TDE induced by NSs for the two channel: binaries (\textit{left panel}, where green cross refer to SN-kick triggered events) and multiples (\textit{right panel}). }
    \label{fig:NS}
\end{figure}

A fraction ($\approx20$\%) of micro-TDE events that occur in the simulated YSCs are triggered by NSs rather than BHs (Fig. \ref{fig:NS}).
The production efficiency of micro-TDEs involving NSs is $(1.4 \pm 0.1) \times 10^{-5}\,\mathrm{M_\odot^{-1}}$, approximately one fifth of the efficiency found for BHs. 
At both $Z$, the largest fraction of events occur in binaries ($\approx 70$\%) and the rest in ($\approx 30$\%) multiples. 
Approximately $91\%$ events occur in \og binaries, while \ex represent only a small fraction of the sample.

No micro-TDE involving NSs is found in the single encounter channel. This naturally arises from the lower mass of NSs and their smaller gravitational focussing, which reduces the cross section for close encounters compared to BHs (Sect. \ref{sec:hyper}).

A significant fraction of binary indued micro-TDEs is actually trigger by SN kick, which account for  $\approx 55$\% of all events.
However, 3D hydrodynamical simulations of newly formed NS–star binaries by \citet{Hirai_2022} show that partial tidal disruptions can occur when the companion star is less massive than the NS, otherwise full disruptions are unlikely to occur. In our simulations, such systems involving low mass stars (See Fig. \ref{fig:NS}) represent only about $\approx 9$\% of the SN-kick subsample.

Most of the stars involved are MS (53\%), followed by a substantial fraction of naked Helium stars (42\%) and a few Giants (4.5\%) and 0.5\% WDs. Star masses ranges between $0.02$ \Ms and $33$ \Ms, while the NSs, as described by the adopted model \citep{fryer12}, span mass ranges between $1$ and $3$ \Ms.

As for BHs, the overall efficiency shows no strong dependence on metallicity or cluster mass. In the case of NSs, the dependence on the cluster central density is also weak (within a factor of two). This behaviour can be understood by considering the interplay between the two dominant channels: while the multiple interaction channel becomes more efficient at higher densities, the kick-induced channel becomes less efficient. The latter requires relatively wide binaries to survive until the SN event, but such systems are more likely to be disrupted in denser environments. For BHs, the kick-induced channel is subdominant by more than an order of magnitude (Figure \ref{fig:eff_ch}), so the overall trend still follows an increase of micro-TDEs with density.\\
In contrast, for NSs, the kick channel accounts for roughly half of all events, effectively balancing the density dependence and resulting in a flatter overall trend.

\section{Discussion}
\label{sec:disc}

\subsection{Comparison with previous works}

A recent study by \citet{kremer2021}, based on Monte Carlo (MC) simulations of YSCs, investigated micro-TDEs of BH-MS stars arising from both single and binary-mediated dynamical encounters. For the single-encounter channel, they predicted a micro-TDE event rate at $z=0$ of the order of $1$–$30$~Gpc$^{-3}$~yr$^{-1}$, depending on the initial cluster radius (either fixed at $r_{\rm h} = 1$~pc or following the \citealt{marks2012} relation). In our work, we estimate a rate from single encounters (see Fig.~\ref{fig:rate}, yellow line) of approximately $10$~Gpc$^{-3}$~yr$^{-1}$, in good agreement with their predictions.

In the same study, they also estimated a significantly higher rate for binary-mediated micro-TDEs, reaching $\approx 20$–$160$~Gpc$^{-3}$~yr$^{-1}$. These events arise exclusively from resonant interactions between dynamically formed BBHs and MS stars, as no \og binaries were included in their models. Their result therefore reflects only the contribution of dynamical BBH-MS encounters and not the full binary-mediated channel.

In contrast, our simulations yield a higher rate for the multiple channel (Fig.~\ref{fig:rate}, blue curve), reaching $\approx 300$–$400$~Gpc$^{-3}$~yr$^{-1}$. The discrepancy with \citet{kremer2021} can be attributed to two main factors: (i) we initialize YSCs with \og binaries, and (ii) our estimate includes a broader range of BH-star configurations (see Sect.~\ref{sec:mult}), beyond the specific BBH+MS interactions considered in their work.

In addition, although our simulations include a variety of stellar types, the vast majority of disrupted stars across the various channels are, as found also by \citet{kremer2021}, MS (90\%), followed by giants (9\%) and WD (1\%). As a result, the different stellar types does not significantly affect the comparison between different rate estimate.\\
Alternative families of stellar clusters have been considered as nurseries of micro-TDEs, such as GCs and OCs.  
\citet{perets2016} and \citet{kremer2019}, using analytical and MC simulations, predicted a micro-TDE rate in GCs of the order of $\approx 10$~Gpc$^{-3}$~yr$^{-1}$.  
\citet{Rastello2018}, through $N$-body simulations, estimated that micro-TDEs involving BBHs in the central regions of OCs could occur at a rate of $\approx 3$-$30$~Gpc$^{-3}$~yr$^{-1}$.  
Although these studies explore different micro-TDE formation channels, employ distinct methodologies, and focus on different stellar environments, our results together with \citep{kremer2021}, suggests that YSCs are likely the dominant producers of micro-TDEs.

For micro-TDEs induced by NSs, assuming a MW star formation rate of SFR $\approx 1$-$2\,\mathrm{M}_\odot \mathrm{yr}^{-1}$ \citep{Chomiuk2011}, we estimate a NS-micro-TDE rate of $\sim$1-$3 \times 10^{-5}\,\mathrm{yr}^{-1}$ per MW-like galaxy. Our results for YSCs are consistent with those reported by \citet{michaely2016}, who predict similar rates ($\sim 1$-$2 \times 10^{-5}\,\mathrm{yr}^{-1}$) depending on the natal kick prescription.

In contrast for GCs, earlier analytical estimates by \citet{perets2016} and \citet{kremer2019} suggested significantly lower NS micro-TDE rates. Specifically, \citet{perets2016} derived a rate of $\sim 3.4-4.8 \times 10^{-7}\,\mathrm{yr}^{-1}$ per MW-like galaxies, roughly an order of magnitude below their corresponding estimate for BH micro-TDEs. \citet{kremer2019} report an even lower rate of $\sim 10^{-8}\,\mathrm{yr}^{-1}$.
GCs are dynamically older systems with lower core densities, resulting in fewer NS-star interactions compared to dense YSCs, where strong dynamical encounters are more frequent.

\subsection{Full vs. partial micro-TDE}

To fully investigate the outcome of close encounters between stars and BHs, accurate hydrodynamical simulations are essential. Indeed, stars may be fully disrupted during the first pericenter passage, or they may undergo partial disruption \citep{perets2016}, either remaining bound to the BH (i.e. forming a BH-star binary) or being completely unbound \citep{wevers2023}. The specific outcome depends on complex physical processes, stellar density profiles and internal stellar structure details, which are beyond the scope of this work.

Our models are designed to capture all the possible channels through which such encounters can bring stars within the tidal disruption radius of COs. As such, the processes that led to micro-TDEs explored in this work can serve as a valuable library of initial conditions for future hydrodynamical simulations, allowing for more targeted investigations with reduced parameter space.

\begin{figure}
 	  \centering
    \includegraphics[width=0.45\textwidth]{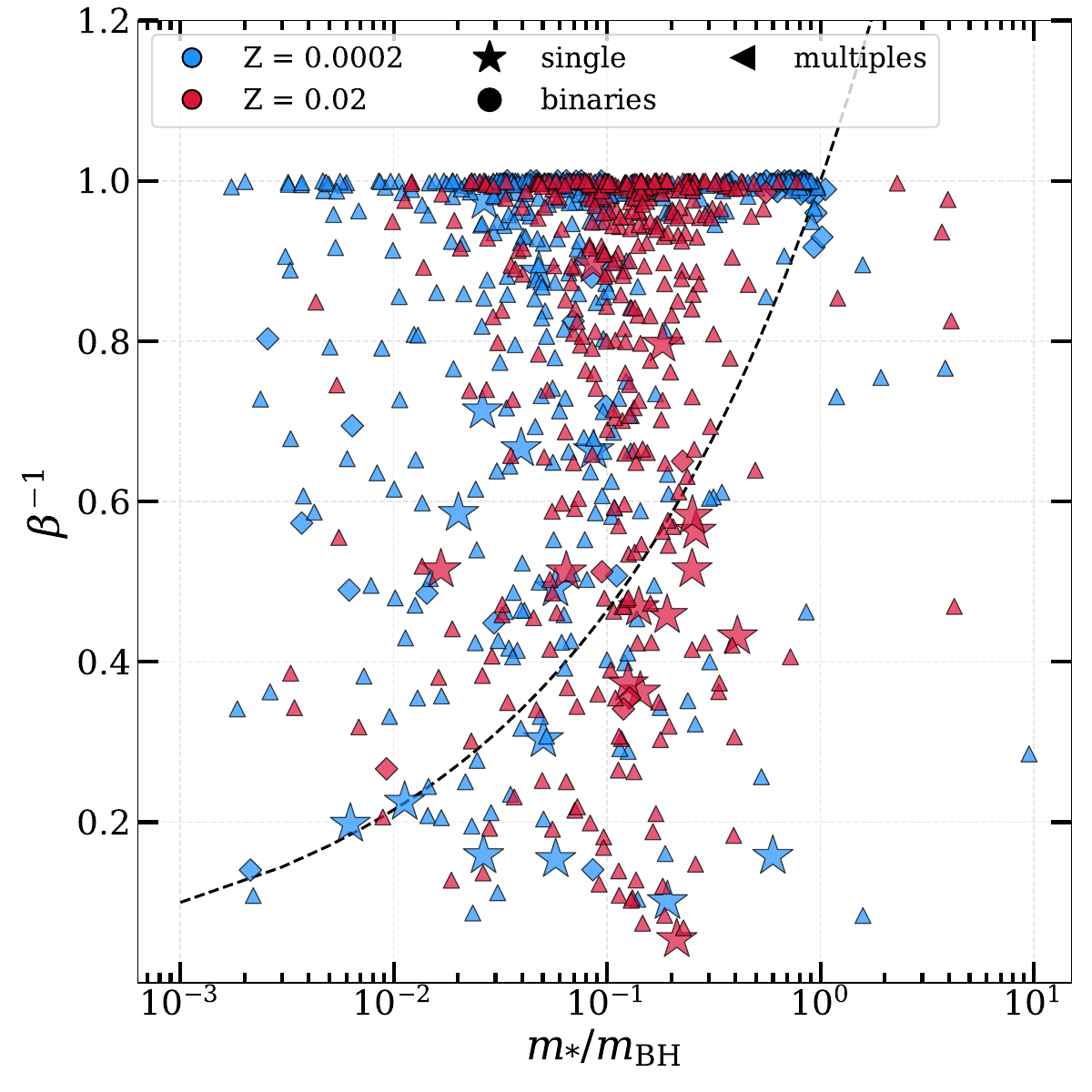}
   \caption{Distribution of micro–TDEs in the plane defined by the stellar-to-BH mass ratio $\left(m_{*}/m_{\rm BH}\right)$ and the inverse of the \be ($\beta^{-1}$). Micro-TDE occurring in our simulated YSCs are shown using the same axes and parameter space as Figure~1 of \citet{kremer_2022}. Each channel (see Section~\ref{sec:channels}) is marked with a different symbol: single (stars), binaries (circles), multiples (triangles). Blue for $Z=0.0002$ and red for $Z=0.02$. The black dashed line indicates the boundary for which ($r_{p} = r_{*}$) where $r_{*}$ is the star radius.
}
    \label{fig:kk22}
\end{figure}

Here, we provide a first-order qualitative estimate of the fraction of full (FTDE) and partial (PTDE) micro-TDE candidates, based on the categories  proposed by \citet{kremer_2022}, who performed hydrodynamical simulations of close encounters between MS stars and mostly equal mass ($m_{\rm{BH}}=10$ \Ms) BHs.  
By comparing our results (Fig. \ref{fig:kk22}) with their Fig.~1, we find that, for both metallicities, approximately 20\% of our candidates fall within the region associated with full disruption (black crosses in their Fig.~1), broadly corresponding to high penetrating encounters ($\beta^{-1} < 0.5$)\footnote{Note that in this subsection and in Fig. \ref{fig:kk22} we use the inverse of the \be, $\beta^{-1}$, to enable direct comparison with the results from \citet[Figure 1]{kremer_2022}}

Among these candidate FTDEs, 85\% originate from multiples, followed by 10\% from single interactions, and binaries (5\%).
The remaining $\approx$80\% of our candidates fall within the region corresponding to PTDEs, which can be either unbound, mostly involving low-mass MS stars ($m_* < 1\,\mathrm{M}_\odot$), or bound, typically associated with more massive stars ($m_* > 1\,\mathrm{M}_\odot$). This outcome is consistent with the results of hydrodynamical simulations by \citet{Vynatheya2024}, who find that PTDEs, occurring at larger pericenter distances (i.e. lower $\beta^{-1}$), are generally more common than FTDEs.

In addition to PTDEs and FTDEs, the process of tidal peeling may also occur, as suggested by the hydrodynamical simulations of \citet{Xin2024}. This mechanism involves low-eccentricity encounters between BHs and stars, during which the star is gradually stripped of its outer layers over multiple orbits, rather than being disrupted in a single passage. Among the micro-TDEs identified in our simulations, those originating from \og{} binaries and involving relatively massive stars ($m_{*} > 1\,M_\odot$) on low-eccentric orbits (Sect \ref{sec:mult}) may represent promising candidates for this process. A more detailed analysis of such cases will be presented in future work.

\subsection{Collision vs. micro-TDE fraction}
\label{sec:coll}

To assess the dynamical relevance of micro-TDEs, we compared their occurrence rate with that of stellar collisions and CO coalescences across clusters of different masses. By combining the simulation outcomes at $Z = 0.0002$ and $Z = 0.02$, we computed the fractions $f_{\rm coll} = N_{\rm micro-TDE}/N_{\rm COLL}$ and $f_{\rm coal} = N_{\rm micro-TDE}/N_{\rm COAL}$ splitting the clusters in four logarithmic mass bins spanning $10^3$-$10^5\,M_\odot$.

We find that the fraction of stellar collisions resulting in TDEs remains nearly constant across all bins, with $f_{\rm coll} \approx 0.004$-$0.008$. Such value is mostly dominated by collisions of \og binaries, while when excluding them, the collision fraction increases up to $\approx 0.1-0.4$. This confirms that micro-TDEs are rare outcomes of stellar dynamical interactions, regardless of the cluster mass. In contrast, the TDE-to-coalescence ratio $f_{\rm coal}$ is significantly larger than unity, ranging from $\approx 5$ to nearly $10$ depending on the mass bin. This suggests that micro-TDEs dominate over CO mergers in number, especially in lower-mass clusters ($M_{\rm{SC}} \leq 10^4\,M_\odot$ where the formation of CO binaries may be dynamically less efficient or delayed.

Assuming the ratios derived from our simulations, we estimate that the detection of roughly $\approx 10$ micro-TDEs would imply the occurrence of roughly $1-2$ CO mergers. 
The combination of micro–TDE detections from upcoming wide-field surveys such as LSST (see Sect.~\ref{sec:detectability}) with the observed rates of CO mergers from GW detectors \citep{abbottGWTC2,gwtc3,Mandel2022} offers a promising channel to constrain the formation channels of GW sources. Since micro–TDEs can serve as tracers of the population of stellar-mass CO in dense SCs, their observed rate provides a valuable prior on the expected number of mergers originating from the same environments.

\subsection{Detectability} \label{sec:detectability}

\begin{figure}
 	  \centering
    \includegraphics[width=0.45\textwidth]{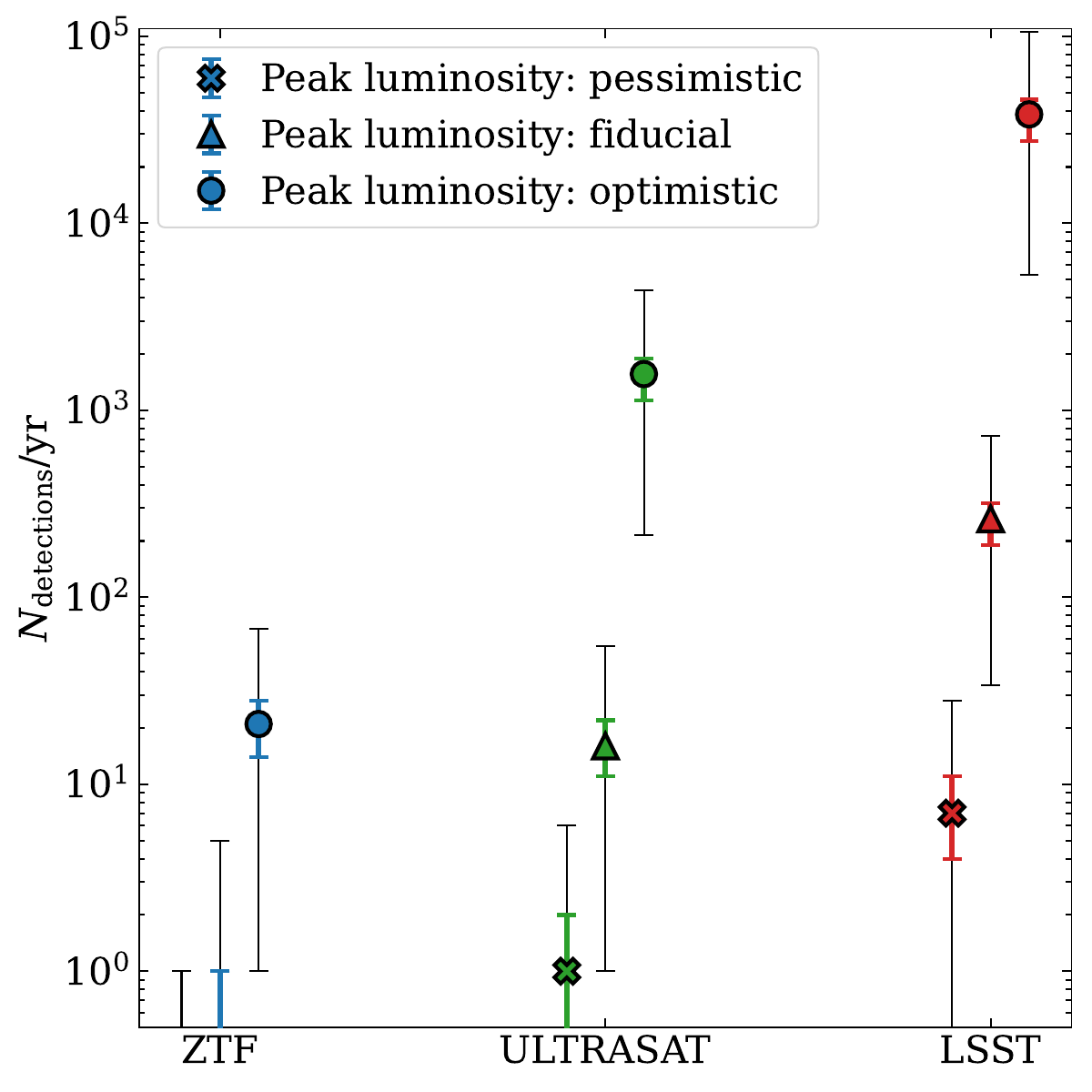}
\caption{Expected number of micro-TDE detections per year for three surveys (ZTF, ULTRASAT, and LSST) under different assumptions for the peak luminosity, based on the wind-reprocessed emission models of \citet{kremer2023} (see main text for details). Symbols indicate the median of the posterior distribution computed using the full simulated sample, while thick colored error bars show the 68\% credible interval. Thin gray error bars represent the range of systematic uncertainties, bracketed by the 0.15th and 99.85th percentiles of the distributions obtained from simulations restricted to low-density and high-density clusters, respectively (see Section~\ref{sec:eta}). Not visible symbols and bars indicate 0 detections.}
    \label{fig:ndetections}
\end{figure}

To assess the detectability of micro-TDE events with current and upcoming optical and near-ultraviolet surveys, we focus on three representative cases: ZTF, LSST for the optical and ULTRASAT for the near-ultraviolet band.

To this end, we combine the predicted event rates (Section~\ref{sec:rates}) with the peak luminosities of micro-TDEs as derived from the wind-reprocessed models of \citet{kremer2023}. In these models, radiation generated by the accretion of disrupted stellar material onto the compact remnant is absorbed and re-emitted as thermal emission at larger radii. The predicted peak luminosities range from \(10^{40}\) to \(10^{43} \ \mathrm{erg\,s^{-1}}\) in the optical (g-band) and near ultraviolet band (NUV), corresponding to absolute magnitudes from  approximately \(-12\) to \(-19\). The peak luminosity is most sensitive to the treatment of disk winds, in particular the parameter "\textit{s}", which determines the fraction of stellar mass ejected in winds versus that accreted \citep[see also][]{Metzger2008,Metzger2022}\footnote{In those papers, the parameter \(s\) is instead denoted as \(p\).}. Following \citet{kremer2023}, we adopt three scenarios to bracket this uncertainty: pessimistic (\(s = 0.2\), lowest peak luminosity), optimistic (\(s = 0.8\), highest peak luminosity), and fiducial (\(s = 0.5\)).

To estimate the band-specific peak luminosities for ZTF, LSST, and ULTRASAT, we use the values from Figure~11 of \citet{kremer2023}, assuming a disrupted stellar mass of \(2 \ \mathrm{M}_\odot\), close to the median in our simulations (see Table~\ref{tab:eta}). Although more massive stars (\(>10 \ \mathrm{M}_\odot\)) yield higher peak luminosities (by up to an order of magnitude), they contribute a small fraction of micro-TDE progenitors (\(<10\%\) at \(Z=0.0002\), and \(<2\%\) at \(Z=0.02\)).

Given the assumed peak luminosities and survey limiting magnitudes, we compute the maximum redshift \(z_\mathrm{max}\) at which a micro-TDE is detectable. The expected number of detections per year is modeled as a Poisson process with mean:
\begin{equation}
\theta = \Gamma(z_\mathrm{max}) \, f_\mathrm{sky} \, \epsilon,
\label{eq:theta_poisson}
\end{equation}
where \(\Gamma(z_\mathrm{max})\) is the cumulative event rate up to \(z_\mathrm{max}\) (Equation~\ref{eq:rate}), \(f_\mathrm{sky}\) is the effective sky fraction covered by the survey with a cadence suitable for detecting micro-TDEs (a few days, based on the light curves in \citealt{kremer2023}), and \(\epsilon\) is the recovery efficiency, i.e., the fraction of detectable events that are both identified and correctly classified as micro-TDEs. For simplicity, we assume \(\epsilon = 1\), although more realistic values likely range between 0.2 and 0.8 \citep{LSST09}.

We draw \(10^4\) samples from the posterior distribution of \(\Gamma(z_\mathrm{max})\) and compute the number of yearly detections by sampling from a Poisson distribution with mean \(\theta\) for each case. This approach incorporates both statistical and systematic uncertainties in the event rates. As with the rate estimates (Figure~\ref{fig:rate}), we repeat the analysis separately for low- and high-density clusters to explore the dependence on cluster properties.
The survey properties (limiting magnitudes, sky fractions) and the summary statistics of \(\Gamma(z_\mathrm{max})\) posteriors are reported in Appendix~\ref{app:rates}.

Figure~\ref{fig:ndetections} summarizes the expected number of micro-TDE detections per year for ZTF, ULTRASAT, and LSST, under three scenarios for the peak luminosity. 
In the optimistic case, all surveys show significantly higher detection rates, with ULTRASAT and LSST predicting up to thousands of events per year. In contrast, the fiducial scenario yields rates nearly an order of magnitude lower, while in the pessimistic case, the detection of micro-TDEs becomes unlikely for both ZTF and ULTRASAT.

ZTF has been operational for nearly six years, during which only one candidate micro-TDE has been reported (ZTF19aailpwl, \citealt{frederik2021}). However, this event is not strictly consistent with the population considered in our study, as it occurred at \( z = 0.35 \) and is associated with an AGN. Excluding this candidate, the observed number of micro-TDEs with ZTF remains zero over six years, a result that aligns with our fiducial and pessimistic models when extrapolated over that timescale.

The large number of expected detections with LSST, even under pessimistic assumptions, highlights its potential to systematically explore the micro-TDE population. The assumed peak luminosity is the dominant source of uncertainty, with detection rate predictions differing by over an order of magnitude across the three scenarios, even after accounting for uncertainties in the intrinsic event rates (Section~\ref{sec:rates}). 

As a result, the number of micro-TDE detections with LSST and ULTRASAT could serve as a powerful diagnostic of the accretion and radiative reprocessing physics in these systems.
If LSST or ULTRASAT detect thousands of events per year, it would be difficult to reconcile this with ZTF’s historical lack of detections, placing constraints on the optimistic scenario. Conversely, a lack of detections in LSST would strongly challenge both our simulation results and the emission models of \citet{kremer2023}. 

A more detailed comparison between observational outcomes and theory would also benefit from incorporating alternative micro-TDE formation environments, such as GCs and AGNs. However, based on our simulations and existing literature \citep{perets2016,kremer2019,kremer2021,kremer2023}, YSCs appear to dominate the micro-TDE production. Thus, we do not expect a substantial shift in detection rates when including other environments.

\subsection{Implications for gravitational wave astronomy}
\label{sec:gw}
TDEs and micro-TDEs are multi-messenger sources \citep{wevers2023}. In addition to EM emission, these transients are expected to produce GWs.
During a TDE, when an "unlucky" star is torn apart, it generates a short burst-like GW signal resulting from the time variation of the mass quadrupole of the BH-star system \citep{Kobayashi2004, East2014, Toscani2020, Pfister2022, Toscani2022,toscani2025}.\\
This GW burst is characterized by an amplitude $h_{\rm{TDE}}$, and a frequency $f_{\rm{TDE}}$:
\begin{align}
    h_{\rm TDE} \approx\ & 2 \times 10^{-22} \, \beta 
    \left( \frac{m_*}{\mathrm{M}_\odot} \right)^{4/3} 
    \left( \frac{m_{\rm BH}}{10^6\,\mathrm{M}_\odot} \right)^{2/3} \nonumber \\
    & \times \left( \frac{r_*}{\mathrm{R}_\odot} \right)^{-1}
    \left( \frac{d}{16\,\mathrm{Mpc}} \right)^{-1}
    \label{eq:hgw}
\end{align}

\begin{equation}
    f_{\rm{TDE}} \approx \beta^{3/2} \times 10^{-4}\,\mathrm{Hz}\,
    \left( \frac{m_*}{\mathrm{M}_\odot} \right)^{1/2}
    \left( \frac{r_*}{\mathrm{R}_\odot} \right)^{-3/2}
\label{eq:fgw}
\end{equation}
where $\beta$ is the \be, $m_*$ and $r_*$ are the mass and radius of the disrupted star, $m_{\rm BH}$ is the mass of the BH and $d$ is the luminosity distance to the source.

\begin{figure}
 	  \centering
    \includegraphics[width=0.45\textwidth]{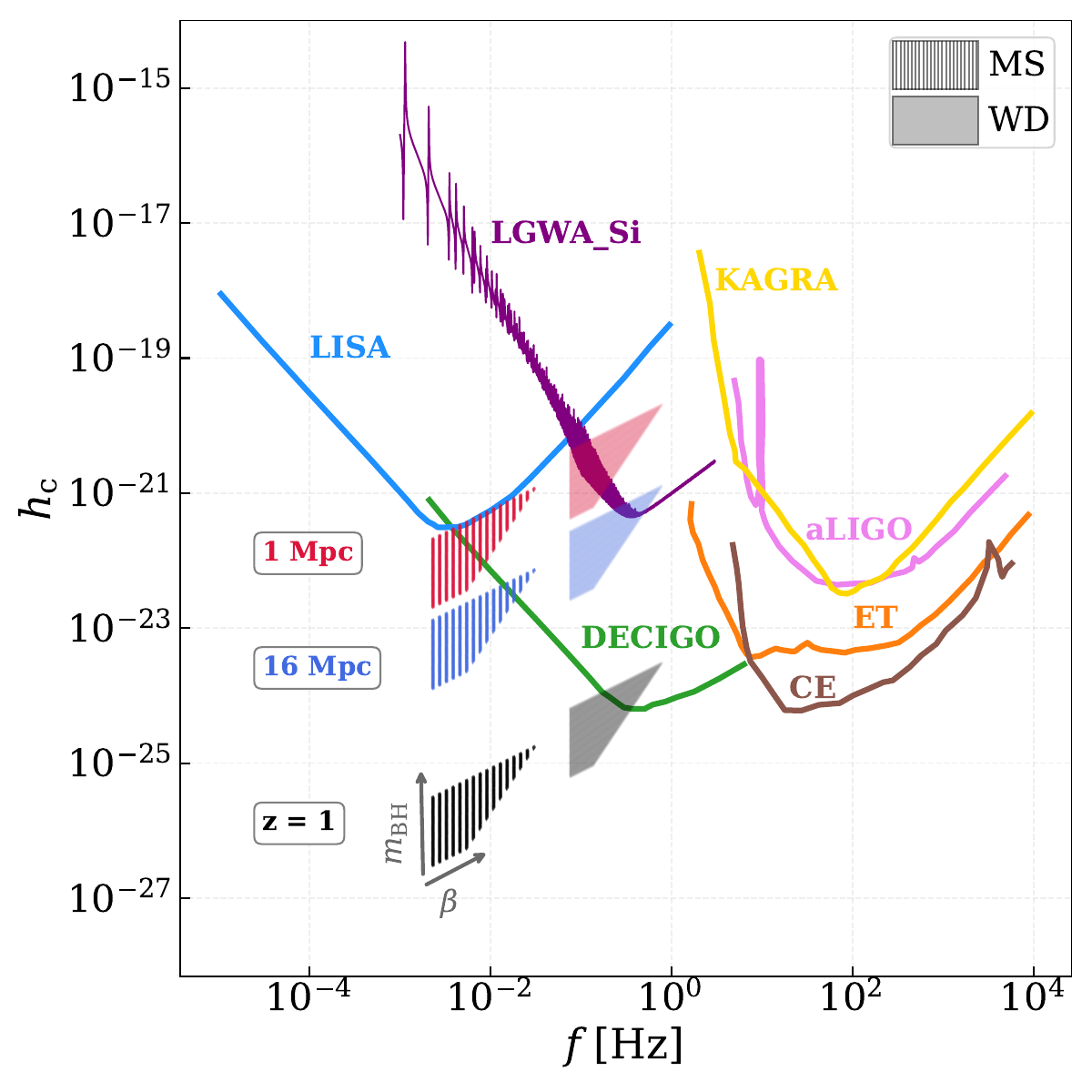}
   \caption{Expected frequency ($f$) and characteristic strain ($h_{c}$) of the GW signal produced by micro-TDEs involving a MS star ($m_{\rm MS} = 0.5\,M_\odot$, hatched bands) and a WD ($m_{\rm WD} = 1.4\,M_\odot$, solid colored bands), shown at three reference distances:$\sim$1 Mpc (red), $\sim$16 Mpc (blue), and $z = 1$ (gray). The two gray arrows in the lower-left corner indicate the directions of increasing BH mass ($m_{\rm BH}$) and \be  ($\beta \equiv r_t / r_p$). These predictions are overlaid with the sensitivity curves of current and future GW detectors: LGWA (purple; \citealt{harms2021}), LISA (sky blue; \citealt{pau2017}), DECIGO (green; \citealt{sato2017}), Kamioka Gravitational Wave Detector (KAGRA, gold; \citealt{abbottGWTC2}), the Einstein Telescope (ET, orange; \citealt{maggiore2020}), Advanced Laser Interferometer Gravitational-Wave Observatory (LIGO, violet; \citealt{aligo}), and the Cosmic Explorer (CE, brown; \citealt{ng2021}).
 }
 \label{fig:res-gw}
\end{figure}

Following the analytical description presented in \citet[][their Eqs.~3-7]{Toscani2020}
\footnote{Following \citep{Toscani2020}, we also use here $\beta_{\rm min} = 1$ (i.e., $r_{\rm p} = r_{\rm t}$) and $\beta_{\rm max} = r_{\rm t}/r_{\rm s}$, where $r_{\rm s}$ is the Schwarzschild radius of the BH.}, we derived the expected GW emission signal of micro-TDEs (Fig. \ref{fig:res-gw}). We consider MS and WD stars disrupted by stellar-mass BHs ($m_{\rm{BH}}$ = 3-100 \Ms). 

The GW signal expected from disrupted WDs falls within the deci-hertz band, and future instruments such as LGWA and DECIGO will be capable of detecting such signals out to large distances, including the most extreme events up to redshift $z=1$.

For MS stars, being less compact than WDs, the GW signal peaks at lower frequencies and is generally fainter, making only nearby events (within 1-16 Mpc) detectable by observatories like DECIGO.

Although micro-TDEs produce fainter GW signals than classical TDEs (mainly because they involve lighter BHs masses, see Eqs.~\ref{eq:hgw} and \ref{eq:fgw}), they remain promising targets for next-generation GW observatories.
Despite their lower amplitude, these events could play a key role in unveiling the presence of otherwise invisible BHs, even in distant galaxies across cosmic time.

In the context of GW detections, previous studies have shown that close encounters between stars and BBHs can perturb the binary orbit and potentially trigger coalescence (e.g., \citealt{ryu_22, Rastello2018, Lopez2019}), suggesting that micro-TDEs could serve as the EM counterparts to GW signals from BBH mergers. 
In this case the GW signal is produced by the BBH merger rather than the stellar disruption itself.
In our simulations we found that small a fraction of events (3\%), limited to metal-poor YSCs, involves BHs that first produce a micro-TDE and later merge with another BH.
However, because of the long delay time between the two events,
none of the cases of merging BBHs recorded in our simulations are directly triggered by the micro-TDEs.

\subsection{Caveats}
\label{sec:cav}

\subsubsection{Relativistic kicks \& Post-Newtonian terms}
\label{sec:rel-kick}
The version of the code \pt used to performed the $N$-body simulations presented in this work, does not account for relativistic recoil kicks. These kicks arise from the asymmetric emission of GWs during the coalescence of BBHs, that imparts a velocity to the remnant BH in order to conserve linear momentum. The magnitude of the recoil depends on the mass ratio and spin configuration of the merging BHs, and can reach several thousand km~s$^{-1}$ \citep{Schnittman2007}.

Such high velocities can easily exceed the escape speed of YSCs, which is typically of the order of $\sim$10–50~km~s$^{-1}$ \citep{Portegies-Zwart10}. As a consequence, BHs that undergo coalescence in YSCs are likely ejected before they can participate in further dynamical interactions or disrupt stars. Only dense and massive environments such as NSCs or AGN disc, where escape velocities are higher \citep{antonini2019}, can retain a significant fraction of post-merger BHs.

To assess the impact of neglecting recoil kicks on our results, we examined the properties of the BHs involved in micro-TDEs in our simulations. Among all micro-TDEs occurring in metal-poor YSCs, 4\% involve a BH that is the remnant of a previous BBH coalescence, where the merger occurred prior to the micro-TDE. For Solar-metallicity YSCs, this fraction drops to 0.8\%. Given these small numbers, we conclude that the absence of recoil kicks in \pt does not significantly affect the main conclusions of this work.

 The code used in this work, \pt, models the orbital decay of coalescing BBHs, due to GW emission using the Peters’ equations \citep{peters1964}. However, the code does not include post-Newtonian corrections for strong, unbound (e.g., hyperbolic) encounters.
While we do not expect major changes in the overall event statistics, a dedicated follow-up analysis using relativistic few-body integrators is planned to assess their potential impact on the dynamics and outcomes of micro-TDEs.

\subsection{Maximum integration time}
\label{sec:time}
The YSCs analysed in this study were integrated with \pt{} up to approximately $\approx 1.5$ Gyr, ensuring that each cluster, given its initial size and mass, experienced at least one relaxation time (see Sec. \ref{sec:ic}). 
Micro-TDEs triggered by SN kicks occur earliest, as expected, with all such events taking place within the first $\sim$10 Myr, immediately following BH formation. 

Events originating from single or binary mediated encounters, as well as from higher-order multiples, exhibit a broader distribution in time. Nevertheless, the majority (about 85\%) of all micro-TDEs occur within the first $\approx 800 $ Myr of SC evolution. A small number of late-time events (up to $\approx$1.1 Gyr) are also present, particularly in more massive YSCs ($M_{\rm{SC}} \approx 10^5$ \Ms).

However the cumulative number of micro-TDEs typically saturates around $\approx 1$ Gyr, indicating that the adopted integration time ($\approx 1.5$ Gyr) is sufficient to capture the effective timescale over which stellar disruptions occur. Simulating the clusters for longer timescales would likely not alter the global picture emerging from our results.

\section{Summary \& conclusions}
\label{sec:conc}

In this paper we have studied the demography of micro-TDEs occurring in metal-poor and metal-rich YSCs. We carried out a suite of direct $N$-body simulations using a customized version of the code \pt in which we include a prescription for TDEs (see Appendix~\ref{app:petar}).

Our cluster models span a range of initial masses, densities, and half-mass radii, designed to cover the observed properties of the MW's YSC population as described by \citet{krumholz2019}. Each cluster is initialized with a King density profile and a mass-dependent primordial binary fraction following \citet{moe17}. The clusters are embedded in the Galactic tidal field and evolved up to 1.5~Gyr.

We identify all events in which a star, either in a close encounter or in a bound orbit, crosses within the tidal disruption radius of a CO, i.e. a BH or  a NS (see Eq.~\ref{eq:cond}), and we distinguish three dynamical channels leading to micro-TDEs: single encounters, binary-mediated interactions including SN-kick trigger, and higher-order multiple systems. For each channel, we estimate the corresponding efficiency and volumetric rate.

The main finding of our works are summarized below:
\begin{itemize}
    \item Micro-TDEs produced by single encounter between stars and BHs are rare at both metallicities with efficiency of the order of  $\eta_{\rm{sing}}\, \approx \,2 \times  10^{-6}$ M$_{\odot}^{-1}$.\\
    \item The binary channel is relatively rare, contributing to only 7\% of the total sample, and exhibits low efficiency, particularly at high metallicity ($\eta_{\rm{bin}} \approx 4 \times 10^{-6}$ M$_\odot^{-1}$; see Sect.~\ref{sec:bin}). It accounts for the majority of \og binaries (about 70\%), where micro-TDEs are primarily driven by stellar evolution processes,such as common envelope (CE) evolution, mass transfer (MT), or supernova kicks, rather than by dynamical interactions. A small fraction of events ($\approx$0.6\%) is directly triggered by a SN explosion.\\
    \item Multiples encounters and interactions in hierarchical triple or quadruples are the most efficient  channel to produce micro-TDEs in YSCs ($\eta_{\rm{mult.}}\, \approx 4.5 \times 10^{-5}$ M$_{\odot}^{-1}$). 
    Half of the triples and quadruples leading to micro-TDEs host at least two BHs or BBHs. 
    \\
    \item Overall, the production efficiency of micro-TDEs is largely independent of metallicity, but increases with the initial density of the cluster due to the higher frequency of dynamical interactions (see Sect.~\ref{sec:eta}).\\
   \item Given the estimated efficiencies, we compute the micro-TDE rate for each evolutionary channel. The highest contribution comes from the multiple encounters channel (Fig.~\ref{fig:rate}, blue curve), with an estimated rate of approximately 300-400 Gpc$^{-3}$ yr$^{-1}$ at $z=0$.
This is followed by the binaries and single channel (Fig.~8, magenta and yellow line), which contributes around 20-10 Gpc$^{-3}$ yr$^{-1}$.
 SN-kick-induced micro-TDEs account for the lowest rate in the range 1-5 Gpc$^{-3}$ yr$^{-1}$.
The total volumetric rate of micro-TDEs is therefore estimated to rise from 350-450 Gpc$^{-3}$ yr$^{-1}$ at $z=0$ up to 2000-3000 Gpc$^{-3}$ yr$^{-1}$ at $z=2$, with the majority of events originating from multiple dynamical encounters.
Accordingly, the number of expected events per year rises from just a few within the local Universe ($z < 0.05$, approximately 200~Mpc) to as many as $10^5$ out to redshift $z = 1$.\\
    \item Among the population of micro-TDEs in YSCs, a small fraction (approximately 20\%) involve NSs. Due to their lower interaction efficiency ($\eta_{\rm NS} \approx 1.4 \times 10^{-5}$~M$_\odot^{-1}$), micro-TDEs involving NSs occur at a reduced rate compared to those involving BHs, of the order of $1$-$3 \times 10^{-5}$~yr$^{-1}$ per MW galaxy.\\
    \item The parameter space of most micro-TDEs derived in YSCs overlaps with the region associated with PTDEs, as identified by \citet{kremer_2022}. However, approximately 20\% of the events produced in YSCs could lead to full disruptions.\\
    \item We assessed the detectability of micro-TDEs by combining our predicted event rates with models for wind-reprocessed emission proposed by \citet{kremer2023}.  
    We find that LSST is the most promising instrument to unveil the micro-TDE population in YSCs.
    Indeed, assuming a peak luminosity of $\sim 10^{40}$-$10^{43}$ erg s$^{-1}$,   LSST  is expected to detect from tens to several tens of thousands of micro-TDEs per year.\\
    
    \item Following the prescription of \citet{Toscani2020}, we estimate the GW emission expected from micro-TDEs. The GW signal is predicted to peak in the deci-Hertz band. Future instruments such as DECIGO and LGWA will be particularly suited to detect these events, especially the closest ones involving the disruption of WDs.
\end{itemize}

The inferred rate of micro-TDEs can provide indirect constraints on the number of stellar collisions in SCs, and can be used to make predictions about the expected population of dormant BHs, CO mergers or other compact remnants.

Micro-TDEs are promising multi-messenger sources that may be detectable by forthcoming EM and GW surveys. Our work represents an extensive study of such transients in YSCs, and the first to systematically explore their formation channels and rates through direct $N$-body simulations over a wide range of initial cluster conditions. To fully capture the complex physical processes involved in these close encounters, future work will require detailed hydrodynamical simulations, ideally incorporating general relativistic effects, in combination with accurate few-body Newtonian dynamical modeling.

\section*{Acknowledgements}
SR gratefully acknowledges helpful discussions with Taeho Ryu, Kyle Kremer, Martina Toscani, Filippo Santoliquido,  Alessandro Alberto Trani, Pavan Vynatheya, Manuel Arca Sedda, Matt Nicholl, Brenna Mockler, Giuseppe Lodato, Daniel Price, Michela Mapelli, Daniel Mar\'{i}n Pina, Helena Ubach, and Kristen Dage.
SR is also thankful to the Virgo and Gaia groups at ICCUB for valuable discussions. SR sincerely thanks Francesca for her constant and invaluable support.
SR further acknowledges financial support from the Beatriu de Pinós postdoctoral fellowship program under the Ministry of Research and Universities of the Government of Catalonia (Grant Reference No. 2021 BP 00213).  
GI is supported by a fellowship grant from la Caixa Foundation (ID 100010434). The fellowship code is LCF/BQ/PI24/12040020.
MG acknowledges financial support from grants PID2021-125485NB-C22, PID2024-155720NB-I00 funded by MCIN/AEI/10.13039/501100011033 and SGR-2021-01069 (AGAUR).
SR, GI and MG acknowledge financial support from grants CEX2019-000918-M, CEX2024-001451-M funded by MICIU/AEI/10.13039/501100011033.
LW thanks the support from the National Natural Science Foundation of China through grant 21BAA00619 and 12233013, the High-level Youth Talent Project (Provincial Financial Allocation) through the grant 2023HYSPT0706, the one-hundred-talent project of Sun Yat-sen University.
This research benefited from scientific interactions during workshops at the Kavli Institute for Theoretical Physics, supported in part by NSF Grant PHY-2309135, and at the Munich Institute for Astro-, Particle and BioPhysics (MIAPbP), which is funded by the Deutsche Forschungsgemeinschaft (DFG, German Research Foundation) under Germany’s Excellence Strategy – EXC-2094 – 390783311.  
We thankfully acknowledge the computing resources provided by the Red Española de Supercomputación (RES) under the projects AECT-2023-3-0015, AECT-2024-2-0037, AECT-2024-3-0017, and AECT-2025-1-0028. This includes access to the Turgalium supercomputer at CETA-CIEMAT, and the AGUSTINA–CESAR cluster at the BIFI Institute (University of Zaragoza). SR and GI acknowledge the use of the NYX supercomputing cluster at ICCUB.

\section*{Data Availability}

The code used for this study will be made publicly available upon publication of the present paper. The simulation data underlying the results presented in this work will be shared upon reasonable request to the corresponding author.

\bibliographystyle{aa_edited} 
\bibliography{biblio} 


\appendix

\section{Updates on \pt and TDE prescriptions} \label{app:petar}  

In the current version of \pt{} \citep{Wang2020b} , TDEs are not included. However, the code does implement a treatment for collisions, which occur when, at the point of closest approach (periastron), the sum of the radii of the two objects exceeds their distance:

\begin{equation}
r_1 \,+ \,r_2 \,> \,r_\mathrm{p}
\label{eq:collision}
\end{equation}

Here, $r_\mathrm{p}$ is the periastron of the two-body orbit, which can be either bound or unbound. When the condition in Equation~\ref{eq:collision} is met, \pt{} passes the system to the stellar and binary evolution module, which determines the outcome of the interaction. Depending on the stellar types involved, the default stellar and binary evolution module \bse{} (see Section~\ref{sec:bse}) may trigger a merger (if both stars are on the MS), a CE phase (if at least one star is evolved with a well-defined core-envelope structure), or the complete destruction of the star (in the case of a collision between a MS star and a COs).

From a practical perspective, TDEs can be seen as a generalization of the stellar collision process. Therefore, we updated the code to include TDEs by introducing an additional condition that triggers a call to the stellar and binary evolution module. Specifically, a TDE is considered to occur if the periastron distance falls within the tidal disruption radius (see Equation~\ref{eq:cond}). TDE interactions are only considered when the two interacting bodies consist of one star (including white dwarfs, WD) and  one dense compact remnant (a BH or a NS). The outcome of the encounter is handled by the stellar and binary evolution module in the same way as for stellar collisions. Since one of the objects is a compact remnant, the TDE of a MS star (either hydrogen-rich or helium-rich) or a WD always results in the complete disruption of the star, with a fraction of its mass being accreted onto the CO (see Section~\ref{sec:bse}). With respect to the original implementation we lower from 50\% to 10\% the amount of star mass accreted on a BH after a star destruction due to a collision or a TDE.

While developing and testing the TDE implementation, we realized that the default treatment of collisions in \pt{} could lead to spurious mergers or CE events. Indeed, in the current setup, whenever the condition in Equation~\ref{eq:collision} is satisfied, the code immediately calls the stellar and binary evolution module, which in turn triggers a merger or CE phase. This occurs regardless of the current physical separation between the two objects and without accounting for potential subsequent dynamical interactions.
As a result, fluctuations in the relative energy and angular momentum of system of two stars, caused by nearby encounters or the influence of the overall cluster and background potential, can satisfy the collision condition in a single time-step, even if the two stars are still far apart or would never actually collide due to later perturbations.

To address this issue, our customized version of \pt{} adds an additional safeguard for both TDEs and collisions: we require that the two objects either already fulfill the collision or TDE condition based on their current separation (i.e., replacing $r_\mathrm{p}$ with the instantaneous distance in Equation~\ref{eq:cond} and Equation~\ref{eq:collision}) or that the pericenter will be reached within the current time-step. In case these conditions are not fulfilled \pt{} will be informed of the time necessary to reach the periastron so that it can account it in handling the close encounter in the next simulation step.
These extra conditions ensure that TDEs identified in our simulations reflect genuine physical encounters rather than transient fluctuations in two-body orbital parameters. 
The algorithm  \ref{alg:tde_merger} presents
the pseudocode of the new updated implementation for collision and TDE.

It is important to note that collisions and TDEs are checked within the part of the code that handles close encounters, binaries, and hierarchical systems. As a result, these conditions can be triggered for pairs of stars both in bound orbits and in unbound orbits, as well as during chaotic few-body interactions. See Section~\ref{sec:channels} for a detailed classification of these cases.

\begin{algorithm}
\caption{Pseudocode of TDE and Merger Detection in the modified \texttt{PeTar} version. This part is included in the method {\it modifyAndInterruptIter} of the class {\it ARInteraction}  in the source file {\it ar\_interaction.hpp}.  }\label{alg:tde_merger}
\begin{algorithmic}[1]
\State \textbf{Input:} Particles $p_1$, $p_2$, binary parameters \texttt{\_bin}, timestep info 
\State Determine object types $t_1$ , $t_2$
\State Compute tidal radius $r_t$ based on star--compact object pair:
\If{$t_1 \in \{\mathrm{Black \ Hole \ (BH)}, \mathrm{Neutron \ Star \ (NS)}\}$ and $t_2 \notin  \{\mathrm{BH},\mathrm{NS}\}$}
    \State $r_t \gets p_2.\texttt{radius} \cdot \left(\frac{p_1.\texttt{mass}}{p_2.\texttt{mass}}\right)^{1/3}$
\ElsIf{$t_2 \in \{\mathrm{BH},\mathrm{NS}\}$ and $t_1 \notin \{\mathrm{BH},\mathrm{NS}\}$}
    \State $r_t \gets p_1.\texttt{radius} \cdot \left(\frac{p_2.\texttt{mass}}{p_1.\texttt{mass}}\right)^{1/3}$
\Else
    \State $r_t \gets -1$
\EndIf
\State Compute pericentre radius $r_p=\texttt{\_bin.semi} (1-\texttt{\_bin.ecc})$
\If{$r_t > 0$ and $r_t > r_{\text{peri}}$} \Comment{Check for possible TDE}
    \State Compute current separation $r$
    \State Compute current relative velocity  $v$
    \If{$r < r_t$}
        \If{\texttt{\_bin.semi} $< 0$}
            \State Trigger \textbf{Hyperbolic TDE} with current $r$
        \ElsIf{\texttt{\_bin.semi} $> 0$}
            \State Trigger \textbf{Binary TDE} with current $r$
        \EndIf
    \ElsIf{$\vec{r} \cdot \vec{v} < 0$} \Comment{Objects are approaching}
        \State Estimate $t_{\text{peri}}$ using orbital parameters:
        \State \hskip1em $M \gets$ mean anomaly
        \State \hskip1em $n \gets$ mean motion
        \State \hskip1em $t_{\text{peri}} \gets M / n$
        \If{$t_{\text{peri}} <$ current time step }
            \If{\texttt{\_bin.semi} $< 0$}
                \State Trigger \textbf{Hyperbolic TDE} at $t_{\text{peri}}$
            \Else
                \State Trigger \textbf{Binary TDE} at $t_{\text{peri}}$
            \EndIf
        \Else
            \State Schedule future collision interrupt at $t_{\text{peri}}$
        \EndIf
    \EndIf
\ElsIf{$r_{\text{collision}} > r_{\text{peri}}$} \Comment{Check for merger}
    \State Compute current separation $r$
    \State Compute current relative velocity  $v$
    \If{$r < r_{\text{collision}}$}
        \If{\texttt{\_bin.semi} $< 0$}
            \State Trigger \textbf{Dynamic Merger} with current $r$
        \Else
            \State Trigger \textbf{Binary Merger} with current $r$
        \EndIf
    \ElsIf{$\vec{r} \cdot \vec{v} < 0$}
        \State Estimate $t_{\text{peri}}$ as above
        \If{$t_{\text{peri}} <$ current time step}
            \If{\texttt{\_bin.semi} $< 0$}
                \State Trigger \textbf{Dynamic Merger} at $t_{\text{peri}}$
            \Else
                \State Trigger \textbf{Binary Merger} at $t_{\text{peri}}$
            \EndIf
        \Else
            \State Schedule future collision interrupt at $t_{\text{peri}}$
        \EndIf
    \EndIf
\EndIf
\end{algorithmic}
\end{algorithm}


\clearpage

\section{Additional Tables}

\subsection{micro-TDE production efficiency}
\label{app:eff}

Table \ref{tab:eta} summarizes the micro-TDE production efficiency per each channel and metallicity as discussed in Sect.\ref{sec:eta}.

\renewcommand{\arraystretch}{1.5}
\begin{table*}[]
\centering
\caption{Summary of the micro-TDE production efficiency, $\eta$, posterior distribution (Equation \ref{eq:posteta}), for different cluster subsets (Section \ref{sec:eta}). The 1$\sigma$, 2$\sigma$, and 3$\sigma$ intervals correspond to the ranges containing 68\%, 95\%, and 99.7\% of the distribution, respectively. The quantities $\langle m_\mathrm{BH} \rangle$ and $\langle m_* \rangle$ indicate the median of mass distribution of the BH and stars involved in the micro-TDE events.}
\label{tab:eta}
\begin{tabular}{lcc|cccc}
\hline
&
&
\multicolumn{1}{l|}{} &
\multicolumn{4}{c}{\begin{tabular}[c]{@{}c@{}}$\eta$\\ ($10^{-5}$ M$^{-1}_\odot$)\end{tabular}} \\ \hline 
&
\multicolumn{1}{c}{\begin{tabular}[c]{@{}c@{}}$\langle m_\mathrm{BH} \rangle$\\ (M$_\odot$)\end{tabular}} & 
\multicolumn{1}{c|}{\begin{tabular}[c]{@{}c@{}}$\langle m_* \rangle$\\ (M$_\odot$)\end{tabular}} & 
Median & 
\multicolumn{1}{c}{1$\sigma$} & 
\multicolumn{1}{c}{2$\sigma$} & 
\multicolumn{1}{c}{3$\sigma$} \\ \hline 
Whole sample  & 19.25 & 1.91 & 5.02 & 4.86--5.20 & 4.70--5.37 & 4.53--5.55 \\ 
Channel Singles  & 22.33 & 1.95 & 0.17 & 0.14--0.20 & 0.11--0.23 & 0.09--0.28 \\ 
Channel Binaries  & 12.83 & 5.90 & 0.37 & 0.32--0.42 & 0.29--0.47 & 0.25--0.52 \\ 
Channel Multiples  & 19.70 & 1.89 & 4.50 & 4.34--4.66 & 4.19--4.82 & 4.03--4.99 \\ 
Channel SN kick  & 4.99 & 15.35 & 0.07 & 0.05--0.10 & 0.04--0.12 & 0.03--0.15 \\ 
Low mass ($M_\mathrm{SC}< 5\mathrm{E}3 \mathrm{M}_\odot$)   & 20.86 & 2.13 & 5.12 & 4.81--5.44 & 4.52--5.77 & 4.23--6.12 \\ 
Intermediate mass ($5\mathrm{E}3 < M_\mathrm{SC}< 5\mathrm{E}4 \mathrm{M}_\odot$)   & 18.17 & 1.89 & 5.75 & 5.50--6.00 & 5.26--6.26 & 5.02--6.53 \\ 
High mass ($M_\mathrm{SC}> 5\mathrm{E}4 \mathrm{M}_\odot$)   & 22.21 & 1.64 & 3.01 & 2.72--3.31 & 2.46--3.63 & 2.20--3.99 \\ 
Low density ($\log \rho = 1$)  & 19.18 & 1.79 & 2.25 & 1.97--2.54 & 1.73--2.85 & 1.50--3.20 \\ 
High density ($\log \rho = 4$)  & 20.07 & 2.01 & 9.16 & 8.49--9.86 & 7.87--10.58 & 7.26--11.36 \\ 
\hline 
 \multicolumn{7}{c}{$Z=0.0002$} \\ \hline 
Whole sample  & 26.51 & 1.58 & 5.08 & 4.85--5.31 & 4.63--5.55 & 4.42--5.80 \\ 
Channel Singles  & 26.86 & 1.03 & 0.17 & 0.13--0.21 & 0.10--0.27 & 0.07--0.33 \\ 
Channel Binaries  & 12.95 & 5.52 & 0.56 & 0.49--0.65 & 0.43--0.73 & 0.36--0.83 \\ 
Channel Multiples  & 27.34 & 1.55 & 4.36 & 4.15--4.58 & 3.95--4.80 & 3.75--5.03 \\ 
Channel SN kick  & 4.81 & 10.87 & 0.08 & 0.06--0.12 & 0.04--0.15 & 0.02--0.20 \\ 
Low mass ($M_\mathrm{SC}< 5\mathrm{E}3 \mathrm{M}_\odot$)   & 25.35 & 1.98 & 6.06 & 5.58--6.56 & 5.15--7.07 & 4.72--7.63 \\ 
Intermediate mass ($5\mathrm{E}3 < M_\mathrm{SC}< 5\mathrm{E}4 \mathrm{M}_\odot$)   & 26.70 & 1.53 & 5.00 & 4.69--5.31 & 4.41--5.63 & 4.13--5.98 \\ 
High mass ($M_\mathrm{SC}> 5\mathrm{E}4 \mathrm{M}_\odot$)   & 27.08 & 1.21 & 3.93 & 3.47--4.43 & 3.06--4.96 & 2.66--5.55 \\ 
Low density ($\log \rho = 1$)  & 23.92 & 1.96 & 2.78 & 2.37--3.23 & 2.01--3.72 & 1.68--4.27 \\ 
High density ($\log \rho = 4$)  & 25.59 & 1.78 & 9.96 & 9.04--10.94 & 8.20--11.95 & 7.39--13.07 \\ 
\hline 
 \multicolumn{7}{c}{$Z=0.02$} \\ \hline 
Whole sample  & 15.70 & 2.12 & 4.97 & 4.72--5.22 & 4.49--5.47 & 4.26--5.75 \\ 
Channel Singles  & 15.09 & 2.54 & 0.17 & 0.13--0.22 & 0.10--0.28 & 0.07--0.35 \\ 
Channel Binaries  & 12.83 & 7.85 & 0.15 & 0.11--0.19 & 0.08--0.25 & 0.05--0.31 \\ 
Channel Multiples  & 15.81 & 2.10 & 4.67 & 4.43--4.91 & 4.21--5.16 & 3.98--5.42 \\ 
Channel SN kick  & 5.13 & 19.92 & 0.07 & 0.05--0.11 & 0.03--0.15 & 0.02--0.20 \\ 
Low mass ($M_\mathrm{SC}< 5\mathrm{E}3 \mathrm{M}_\odot$)   & 16.97 & 2.31 & 4.20 & 3.81--4.62 & 3.45--5.05 & 3.10--5.53 \\ 
Intermediate mass ($5\mathrm{E}3 < M_\mathrm{SC}< 5\mathrm{E}4 \mathrm{M}_\odot$)   & 15.56 & 2.10 & 6.79 & 6.38--7.22 & 5.99--7.66 & 5.61--8.13 \\ 
High mass ($M_\mathrm{SC}> 5\mathrm{E}4 \mathrm{M}_\odot$)   & 15.08 & 2.09 & 2.13 & 1.79--2.50 & 1.50--2.90 & 1.24--3.36 \\ 
Low density ($\log \rho = 1$)  & 15.85 & 1.79 & 1.69 & 1.36--2.08 & 1.08--2.51 & 0.83--3.01 \\ 
High density ($\log \rho = 4$)  & 16.34 & 2.16 & 8.18 & 7.23--9.21 & 6.38--10.28 & 5.57--11.49 \\ 
\hline 
\end{tabular} 
\end{table*}

\renewcommand{\arraystretch}{1.5}
\newpage

\subsection{micro-TDE  Rates for ZTF, ULTRASAT and LSST} \label{app:rates}

Table \ref{tab:detection} summarises the properties and the posterior distribution of the micro-TDE rates (events per year) for the three surveys analysed in this work (Section \ref{sec:detectability}): ZTF (Zwicky Transient Facility at Palomar Observatory; \citealt{Bellm2019}), LSST (Legacy Survey of Space and Time at the Vera C. Rubin Observatory; \citealt{Ivezic2019}), and ULTRASAT (a space-based NUV imaging mission; \citealt{Sagiv2014,Shvartzvald2024}).

For ZTF, we considered the Northern-equatorial sky survey with a reported  mean limit magnitude of 20.5 in the g-band and a sky coverage of $15000 \ \mathrm{deg}^2$ with 3 days cadence ($f_\mathrm{sky}=0.364$). 
For LSST, we considered the Wide Fast Deep survey with a mean limit magnitude of 25.0 in the g-band and a sky coverage of $19600 \ \mathrm{deg}^2$ with 2-4 days cadence ($f_\mathrm{sky}=0.364$). 
For ULTRASAT, we considered the low cadence survey with a sky coverage of $6800 \ \mathrm{deg}^2$ ($f_\mathrm{sky}=0.165$) and expected limit magnitude of 22.5 in the near ultraviolet band.

The rates and the sky fraction have been used in Section \ref{sec:detectability} (Equation \ref{eq:theta_poisson}) to estimate the number of expected detection of micro-TDEs for each one of the three surveys.

\renewcommand{\arraystretch}{1.8}
\begin{table*}[]
\centering
\caption{Summary of the expected detection rate, $\Gamma$, for three different instruments/surveys (with their observing bands indicated in parentheses), based on their maximum redshift $z_\mathrm{max}$ at which events with peak luminosity $M_\mathrm{peak}$ are detectable.
The number of expected detections per year follows a Poisson distribution with mean given by Equation~\ref{eq:theta_poisson}, which depends on the detection rate, sky coverage ($f_\mathrm{sky}$), and detection efficiency.
 The reported value corresponds to the median of the expected distribution, with subscript and superscript denoting the central 68\% confidence interval, and the range in square brackets indicating the 99.7\% confidence interval. The fiducial, low density and high density columns refer to the results obtained assuming the correspondent production efficiency in Table \ref{tab:eta}. The three peak luminosity scenarios refer to the models presented in \cite{kremer2023} for different mass accretion efficiency on the BH (pessimistic: low efficiency, optimistic: high efficiency).}
\label{tab:detection}
\begin{tabular}{lcccc|ccc}
\hline
Instrument & M$_\mathrm{max}$ & M$_\mathrm{peak}$ & $z_\mathrm{max}$ & $f_\mathrm{sky}$  &
\multicolumn{1}{c}{\begin{tabular}[c]{@{}c@{}}$\Gamma(z_\mathrm{max})$ fiducial \\ ($\mathrm{yr}^{-1}$)\end{tabular}} &
\multicolumn{1}{c}{\begin{tabular}[c]{@{}c@{}}$\Gamma(z_\mathrm{max})$ low density \\ ($\mathrm{yr}^{-1}$)\end{tabular}} &
\multicolumn{1}{c}{\begin{tabular}[c]{@{}c@{}}$\Gamma(z_\mathrm{max})$ high density \\ ($\mathrm{yr}^{-1}$)\end{tabular}} \\
\hline \multicolumn{8}{c}{Peak luminosity: fiducial} \\ \hline 
 ZTF & 20.8 (g-band) &  -14.24 & 0.023 &  0.360 & \begin{tabular}[c]{@{}c@{}}$1.58^{1.91}_{1.13}$\\ {[}0.36--2.27{]}\end{tabular}  & \begin{tabular}[c]{@{}c@{}}$0.52^{0.69}_{0.35}$\\ {[}0.09--1.14{]}\end{tabular}  & \begin{tabular}[c]{@{}c@{}}$2.58^{3.22}_{1.86}$\\ {[}0.58--4.34{]}\end{tabular} \\ 
 ULTRASAT & 22.5 (NUV) &  -15.65 & 0.090 &  0.170 & \begin{tabular}[c]{@{}c@{}}$103.62^{125.17}_{74.06}$\\ {[}23.52--148.70{]}\end{tabular}  & \begin{tabular}[c]{@{}c@{}}$34.16^{45.76}_{23.53}$\\ {[}7.01--74.64{]}\end{tabular}  & \begin{tabular}[c]{@{}c@{}}$167.83^{210.40}_{118.79}$\\ {[}32.37--288.78{]}\end{tabular} \\ 
 LSST & 25.0 (g-band) &  -14.44 & 0.157 &  0.480 & \begin{tabular}[c]{@{}c@{}}$554.08^{666.50}_{403.79}$\\ {[}117.33--790.79{]}\end{tabular}  & \begin{tabular}[c]{@{}c@{}}$181.55^{243.86}_{125.24}$\\ {[}36.11--374.94{]}\end{tabular}  & \begin{tabular}[c]{@{}c@{}}$899.97^{1121.91}_{643.10}$\\ {[}219.49--1517.28{]}\end{tabular} \\ 
\hline \multicolumn{8}{c}{Peak luminosity: pessimistic} \\ \hline 
 ZTF & 20.8 (g-band) &  -11.74 & 0.007 &  0.360 & \begin{tabular}[c]{@{}c@{}}$0.05^{0.06}_{0.04}$\\ {[}0.01--0.07{]}\end{tabular}  & \begin{tabular}[c]{@{}c@{}}$0.02^{0.02}_{0.01}$\\ {[}0.00--0.04{]}\end{tabular}  & \begin{tabular}[c]{@{}c@{}}$0.08^{0.10}_{0.06}$\\ {[}0.02--0.14{]}\end{tabular} \\ 
 ULTRASAT & 22.5 (NUV) &  -13.30 & 0.032 &  0.170 & \begin{tabular}[c]{@{}c@{}}$4.44^{5.36}_{3.20}$\\ {[}1.07--6.37{]}\end{tabular}  & \begin{tabular}[c]{@{}c@{}}$1.46^{1.97}_{1.00}$\\ {[}0.30--3.13{]}\end{tabular}  & \begin{tabular}[c]{@{}c@{}}$7.23^{9.05}_{5.20}$\\ {[}1.44--12.23{]}\end{tabular} \\ 
 LSST & 25.0 (g-band) &  -11.74 & 0.049 &  0.480 & \begin{tabular}[c]{@{}c@{}}$15.86^{19.10}_{11.47}$\\ {[}3.35--22.81{]}\end{tabular}  & \begin{tabular}[c]{@{}c@{}}$5.26^{7.00}_{3.66}$\\ {[}1.08--11.17{]}\end{tabular}  & \begin{tabular}[c]{@{}c@{}}$25.75^{32.13}_{18.21}$\\ {[}5.65--43.50{]}\end{tabular} \\ 
\hline \multicolumn{8}{c}{Peak luminosity: optimistic} \\ \hline 
 ZTF & 20.8 (g-band) &  -16.94 & 0.075 &  0.360 & \begin{tabular}[c]{@{}c@{}}$60.19^{72.32}_{43.65}$\\ {[}15.59--86.37{]}\end{tabular}  & \begin{tabular}[c]{@{}c@{}}$19.81^{26.54}_{13.51}$\\ {[}4.35--42.47{]}\end{tabular}  & \begin{tabular}[c]{@{}c@{}}$97.52^{122.51}_{70.10}$\\ {[}22.16--165.06{]}\end{tabular} \\ 
 ULTRASAT & 22.5 (NUV) &  -19.24 & 0.399 &  0.170 & \begin{tabular}[c]{@{}c@{}}$9476.41^{11421.41}_{6832.95}$\\ {[}2213.20--13591.51{]}\end{tabular}  & \begin{tabular}[c]{@{}c@{}}$3133.84^{4176.86}_{2132.37}$\\ {[}669.56--6564.86{]}\end{tabular}  & \begin{tabular}[c]{@{}c@{}}$15410.92^{19206.03}_{10947.78}$\\ {[}3246.85--26417.07{]}\end{tabular} \\ 
 LSST & 25.0 (g-band) &  -18.63 & 0.819 &  0.480 & \begin{tabular}[c]{@{}c@{}}$80441.90^{96546.25}_{57954.33}$\\ {[}18729.98--115310.51{]}\end{tabular}  & \begin{tabular}[c]{@{}c@{}}$26226.62^{35557.07}_{18010.33}$\\ {[}5269.67--56786.63{]}\end{tabular}  & \begin{tabular}[c]{@{}c@{}}$130614.00^{162814.77}_{92870.77}$\\ {[}27930.43--223020.88{]}\end{tabular} \\ 
\hline 
\end{tabular} 
\end{table*} 
\renewcommand{\arraystretch}{1.8}

\end{document}